\documentclass[3p]{elsarticle}
\biboptions{sort&compress}

\usepackage{amssymb}
\usepackage[utf8]{inputenc}
\usepackage[english]{babel}
\usepackage[T1]{fontenc}
\usepackage{amsmath}
\usepackage{amsthm}
\usepackage{hyperref}
\usepackage{caption}
\usepackage{subcaption}
\usepackage[linesnumbered,ruled,vlined]{algorithm2e}
\usepackage{tikz}
\usepackage{quantikz}
\usepackage{physics}
\usepackage[normalem]{ulem}
\usepackage{listings}
\usepackage{multirow}
\usepackage{xcolor}
\usepackage{comment}
\usepackage{mathdots}
\usepackage{units}
\usepackage{floatrow}

\hypersetup{pdfauthor=author}

\newtheorem{theorem}{Theorem}
\newtheorem{corollary}{Corollary}[theorem]

\theoremstyle{definition}

\newcounter{psdcode}
\newenvironment{pseudocode}
  {\refstepcounter{psdcode}%
    \begin{algorithm}}
  {\end{algorithm}\addtocounter{algocf}{-1}}

\SetKwRepeat{Struct}{struct \{}{\}}
\SetKwRepeat{Class}{class \{}{\}}

\newcommand{\length}{\textnormal{length}}
\newcommand{\Null}{\textnormal{null}}
\SetKw{KwStep}{step}
\SetKwFunction{measure}{measure}
\SetKwFunction{statedecomposition}{state\_decomposition}
\SetKwFunction{sparsestatedecomposition}{sparse\_state\_decomposition}
\SetKwFunction{topdowntreewalk}{top\_down\_tree\_walk}
\SetKwFunction{inittopdown}{initialize\_top\_down}
\SetKwFunction{angletree}{angle\_tree}
\SetKwFunction{isleaf}{is\_leaf}
\SetKwFunction{leftview}{left\_view}
\SetKwFunction{cswaps}{cswaps}
\SetKwFunction{bottomuptreewalk}{bottom\_up\_tree\_walk}
\SetKwFunction{initbottomup}{initialize\_bottom\_up}
\SetKwFunction{initbidirectional}{initialize\_bidirectional}
\SetKwFunction{initsparsebidirectional}{initialize\_sparse\_bidirectional}
\SetKwFunction{addqubit}{add\_qubit}

\journal{an international journal}

\begin{document}

\begin{frontmatter}

\title{Configurable sublinear circuits for quantum state preparation}

\author[1]{\href{https://orcid.org/0000-0002-0308-8701}{Israel F. Araujo}\corref{author}}
\author[2,3]{\href{https://orcid.org/0000-0002-3177-4143}{Daniel K. Park}}
\author[1]{\href{https://orcid.org/0000-0002-8980-6742}{Teresa B. Ludermir}}
\author[4]{\href{https://orcid.org/0000-0002-3261-8265}{Wilson R. Oliveira}}
\author[5,6]{\href{https://orcid.org/0000-0002-8604-0913}{Francesco Petruccione}}
\author[1]{\href{https://orcid.org/0000-0003-0019-7694}{Adenilton J. da Silva}}

\cortext[author]{Corresponding author.\\\textit{E-mail address:} ifa@cin.ufpe.br}
\address[1]{Centro de Inform\'atica, Universidade Federal de Pernambuco, 50740-560, Recife, Pernambuco, Brazil}
\address[2]{Department of Applied Statistics, Yonsei University, Seoul, Republic of Korea}
\address[3]{Department of Statistics and Data Science, Yonsei University, Seoul, Republic of Korea}
\address[4]{Departamento de Estat\'istica e Inform\'atica, Universidade Federal Rural de Pernambuco, Recife, Pernambuco, Brazil}
\address[5]{Quantum Research Group, School of Chemistry and Physics, University of KwaZulu-Natal, Durban, 4001, South Africa}
\address[6]{National Institute for Theoretical and Computational Sciences (NITheCS), 4001, South Africa}

\begin{abstract}
The theory of quantum algorithms promises unprecedented benefits of harnessing the laws of quantum mechanics for solving certain computational problems. A persistent obstacle to using such algorithms for solving a wide range of real-world problems is the cost of loading classical data to a quantum state. Several quantum circuit-based methods have been proposed for encoding classical data as probability amplitudes of a quantum state. However, they require either quantum circuit depth or width to grow linearly with the data size, even though the other dimension of the quantum circuit grows logarithmically. In this paper, we present a configurable bidirectional procedure that addresses this problem by tailoring the resource trade-off between quantum circuit width and depth. In particular, we show a configuration that encodes an $N$-dimensional state by a quantum circuit with $O(\sqrt{N})$ width and depth and entangled information in ancillary qubits. We show a proof-of-principle on five quantum computers.
\end{abstract}


\end{frontmatter}

\section{Introduction}

Quantum algorithms assume an initial quantum state prepared before the computation.
The worst case complexity of preparing an arbitrary quantum state is exponential with the number of qubits~\cite{shende2006synthesis}.
For this reason, the most significant quantum speed-ups occur when the quantum algorithm~\cite{deutsch_rapid_1992, HOGG19961, grover_fast_1996, simon_power_1997, terhal_single_1998, shor_polynomial-time_1999} operates on an input state that is easy to prepare, such as the uniform superposition of all computational basis states. For algorithms that rely on loading data into an arbitrary quantum state, an efficient means to prepare input states is a prerequisite to quantum speed-ups~\cite{Aaronson_nature2020, biamonte2017quantum, harrow2009quantum,PhysRevLett.127.060503}.

While the quantum state preparation models based on quantum oracles or quantum random access memory are useful for evaluating the lower bounds of the computational cost and identifying the complexity class, implementations of them must be considered in practice. In particular, the quantum speed-up can vanish without an efficient implementation of quantum state preparation when quantum algorithms carry classical data in a non-uniform quantum superposition. Examples of such instances include Quantum Machine Learning (QML)~\cite{lloyd2013quantum,biamonte2017quantum, stoudenmire_supervised_2017, schuld_implementing_2017, schuld_supervised_2019, benedetti_parameterized_2019, levine_quantum_2019, Aaronson_nature2020, blank_quantum_2020}, Quantum Memories (QMem)~\cite{Trugenberger_2001, ventura_quantum_2000, trugenberger_quantum_2002,PhysRevLett.100.160501, silva_weightless_2010, DEPAULANETO2019101,park2019circuit}, and Quantum Linear Algebra (QLA)~\cite{harrow2009quantum, lloyd_quantum_2014, Childs_2017, biamonte2017quantum, wossnig2018quantum, Rebentrost_2018}.
Quantum machine learning algorithms try to estimate a target function from a finite set of example points by unveiling correlations between inputs and outputs of the correspondent function~\cite{mitchell_machine_2013, biamonte2017quantum, schuld_supervised_2019}. Quantum memories must store a set of samples from a configuration space as a superposition state before the information is retrieved using the algorithm~\cite{Trugenberger_2001}. Quantum linear algebra algorithms operate with a critical assumption that classical data has been efficiently encoded as probability amplitudes of a quantum state without which the quantum speed-up vanishes~\cite{harrow2009quantum, Aaronson_nature2020, biamonte2017quantum,PhysRevLett.127.060503}. All of the above emphasizes the importance of developing efficient quantum state preparation algorithms for broad application of quantum computing techniques on classical data.

\begin{figure}[t]
  \centering
\begin{subfigure}[b]{.49\textwidth}
  \centering
\begin{tikzpicture}[level distance=1cm,
    level 1/.style={sibling distance=3.6cm, level distance=1cm},
    level 2/.style={sibling distance=1.8cm, level  distance=1cm},
    level 3/.style={sibling distance=.9cm, level  distance=1.0cm},
    level 4/.style={sibling distance=.9cm, level  distance=1.0cm},
    circle1/.style={circle, draw=black!60, fill=white!5, very thick, minimum size=8mm},
    circle2/.style={circle, draw=black!60,dashed, fill=black!5, very thick, minimum size=8mm}]
    \node[circle1]{$\eta_{1,3}$}
        child{
        	node[circle1]{$\eta_{1,2}$}
            child {
                node[circle1] {$\eta_{1,1}$}
                child {
                    node[circle2] {$x_0$}
            	}
                child {
                    node[circle2] {$x_1$}
                }
        	}
            child {
                node[circle1] {$\eta_{2,1}$}
                child {
                    node[circle2] {$x_2$}
            	}
                child {
                    node[circle2] {$x_3$}
                }
            }
        }
        child{
        node[circle1]{$\eta_{2,2}$}
            child {
                node[circle1] {$\eta_{3,1}$}
                child {
                    node[circle2] {$x_4$}
            	}
                child {
                    node[circle2] {$x_5$}
                }
            }
            child {
                node[circle1] {$\eta_{4,1}$}
                child {
                    node[circle2] {$x_6$}
            	}
                child {
                    node[circle2] {$x_7$}
                }
            }
        };
\end{tikzpicture}
    \caption{}
    \label{fig:statetree1}
\end{subfigure}
\begin{subfigure}[b]{.49\textwidth}
  \centering
\begin{tikzpicture}[level distance=1cm,
    level 1/.style={sibling distance=3.6cm, level distance=1cm},
    level 2/.style={sibling distance=1.8cm, level  distance=1cm},
    level 3/.style={sibling distance=.9cm, level  distance=1.0cm},
    circle1/.style={circle, draw=black!60, fill=white!5, very thick, minimum size=7mm}]
    \node[circle1]{$\alpha_{1,3}$}
        child{
        	node[circle1]{$\alpha_{1,2}$}
            child {
                node[circle1] {$\alpha_{1,1}$}
        	}
            child {
                node[circle1] {$\alpha_{2,1}$}
            }
        }
        child{
        node[circle1]{$\alpha_{2,2}$}
            child {
                node[circle1] {$\alpha_{3,1}$}
            }
            child {
                node[circle1] {$\alpha_{4,1}$}
            }
        };
\end{tikzpicture}
    \caption{}
    \label{fig:angletree1}
\end{subfigure}

\caption{Tree representations of quantum state preparation algorithms. (a) State decomposition tree generated by Algorithm~\ref{alg:state_decomposition2} with an 8-dimensional input vector $\mathbf{x}$ (dashed nodes). The complex argument terms $\Omega_{i,k}$ were omitted for readability. (b) Angle tree generated by Algorithm~\ref{alg:angle_tree2} with an 8-dimensional input vector. The correspondent phase angles $\lambda_{j,v}$ were omitted for readability.}
\end{figure}
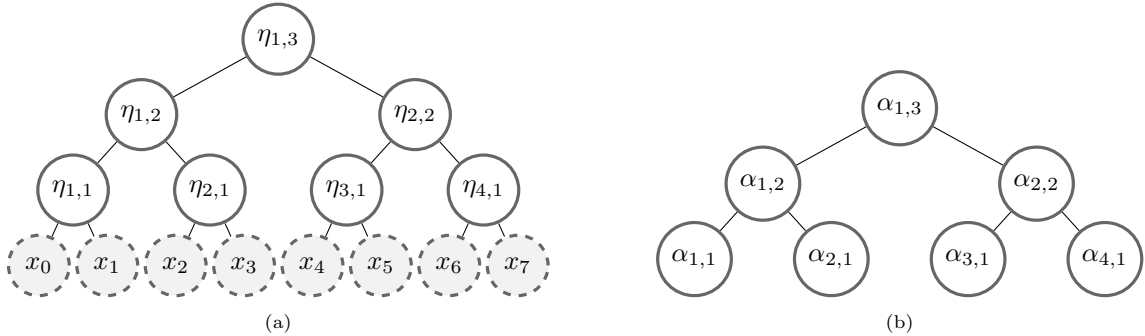

Several solutions to the problem of quantum state preparation have been proposed~\cite{ventura_initializing_1999, grover_synthesis_2000, long_efficient_2001, mottonen_transformation_2005, shende2006synthesis, plesch_quantum-state_2011,cortese2018loading, araujo_divide-and-conquer_2021}, but all produce circuits with width or depth growing at least linearly with the size of the input vector~\cite{shende2006synthesis}. 
For example, the top-down method proposed in Ref.~\cite{ventura_initializing_1999} achieves the exponential compression of the quantum circuit width while requiring $O(N)$ quantum circuit depth for $N$-dimensional data. On the other extreme end, the bottom-up method \cite{araujo_divide-and-conquer_2021} achieves the exponential compression of the quantum circuit depth while requiring  $O(N)$ quantum circuit width and entangled information in ancillary qubits. Since there is an extra resource overhead in many quantum algorithms due to the quantum measurement postulate~\cite{park2019circuit,Park2019NJP}, such linear cost can impose restrictions on possible speed-ups, dominating the computational cost of the intended quantum application. Other approaches have reduced circuit complexity to initialize an approximate quantum state \cite{grover_synthesis_2000,low2018trading,zoufal2019quantum,Kuzmin2020variationalquantum}, but this paper targets the exact state preparation with entangled ancillary qubits.

This work presents an quantum state preparation method that achieves sublinear scaling on both quantum circuit resources. More specifically, a bidirectional strategy that effectively combines the aforementioned approaches in a way that the trade-off between computational time and space can be configured. Both temporal and spatial complexities depend on the parameter $s \in [1..n]$, which adjusts the trade-off between computational time and space. Given an $N$-dimensional input vector, the total time complexity of the bidirectional algorithm is $O_c(N)+O_d(2^s+\log_2^2(N)-s^2)$, where $O_c(N)$ is the time of the classical preprocessing to create the quantum circuit and $O_d(2^s+\log_2^2(N)-s^2)$ is the quantum circuit depth. Typically the same input vector is loaded $l \gg N$ times, and hence the amortized computational time is $O_d(2^s+\log_2^2(N)-s^2)$. Note that classical preprocessing is also common in classical computing and is necessary in other quantum state preparation methods as well. The spatial complexity (i.e. the width) of the circuit is $O_w((s+1)\nicefrac{N}{2^s})$.

Besides the sublinear circuit cost, the ability to customize the exchange between these quantum resources is advantageous when realistic quantum hardware specifications are considered as one resource can be cheaper than the other to scale up. For instance, it is a useful feature for future Noisy Intermediate-Scale Quantum (NISQ) devices with the promise of computers with a large number of physical qubits
~\cite{noauthor_ibms_2020}, albeit noise limits the depth of the circuits~\cite{preskill2018quantum}.

This paper is divided into four sections. Section \ref{sec:review} reviews two strategies for loading classical information into quantum devices, namely top-down~\cite{mottonen_transformation_2005} and bottom-up~\cite{araujo_divide-and-conquer_2021} approaches. The former is used by quantum computing libraries~\cite{gadi_aleksandrowicz_2019_2562111, bergholm_pennylane_2020} as the method for general quantum amplitude initialization. These methods are at the two opposite ends of the quantum circuit cost spectrum requiring either the maximal circuit depth or width to minimize the other resource. Section \ref{sec:bidirectional} presents the main result, a bidirectional method that combines the top-down and bottom-up strategies in a configurable manner. Complexity expressions for the bidirectional method are established in Section~\ref{sec:complexity}, which shows that the bottom-up and the top-down strategies are recovered when $s=1$ and $s=n$, respectively, and that sublinear scaling on both depth and width is possible when $s=\nicefrac{n}{2}$. Proof-of-principle experiments performed on superconducting and ion-trap based quantum devices are presented in Section~\ref{sec:experiment}. Section~\ref{sec:conclusion} presents the conclusion and perspectives for future work.

\section{Quantum state preparation with linear cost}
\label{sec:review}

\subsection{Tree representation}
\label{sec:tree}

Quantum state preparation algorithms aim to create a state $\sum_{p}|x_p|e^{i\omega_p}\ket{p}$ that encodes a normalized vector $\mathbf{x}=(|x_0|e^{i\omega_0}, \dots, |x_{N-1}|e^{i\omega_{N-1}})$ as the probability amplitudes. Several of the existing methods can be understood as a walk on a binary tree \cite{mottonen_transformation_2005,bergholm_quantum_2005,shende2006synthesis,cortese2018loading,araujo_divide-and-conquer_2021}. Each tree node corresponds to a controlled gate operation and the height increases with the number of qubits (see Fig.~\ref{fig:statetree1} and Fig.~\ref{fig:angletree1}). Two edges stemming from each node indicate that each controlled gate operation splits the Hilbert space into two subspaces. Therefore, after $n$ layers, there can be $2^n$ subspaces with distinct probability amplitudes. Depending on the choice of the walk direction, different state preparation strategies, such as top-down and bottom-up approaches, can be constructed.

To explain the procedure, four parameters~\cite{mottonen_transformation_2005} defined by the target vector $\mathbf{x}$ are introduced as
\begin{align} 
\Omega_{i,k} &= \sum_{l=0}^{2^{k}-1} \omega_{(i-1)2^{k}+l}/2^{k-1} \label{eq:Omega} \\
\eta_{i,k} &= \sqrt{\sum_{l=0}^{2^{k}-1}|x_{(i-1)2^k+l}|^2} \label{eq:eta} \\
\lambda_{j,v} &=  \Omega_{2j,v-1} - \Omega_{j,v} \label{eq:lambda} \\
\beta_{j,v} &= \eta_{2j,v-1} / \eta_{j,v} \label{eq:beta}
\end{align}
where $j=1,2,\dots,2^{n-v}$, $v=1,2,\dots,n$, and $n=\log_2(N)$.
These parameters are used to construct the tree representations of the state preparation algorithms, namely the state tree (Fig.~\ref{fig:statetree1}) and the angle tree (Fig.~\ref{fig:angletree1}). Indices $k$ and $v$ indicate a tree level in ascending order from the leaf nodes to the root, $i$ and $j$ are node indices at a given level. The nodes of these trees are complex values that represent the amplitudes of the quantum state to be encoded and the rotation angles for the construction of the encoding quantum circuit. The magnitude and complex argument of the state tree amplitudes are obtained through $\eta_{i,k}$ and $\Omega_{i,k}$, respectively. When $k=0$, the parameters point to the input vector $\mathbf{x}$. Equations \eqref{eq:lambda} and \eqref{eq:beta} determine rotation values of the angle tree nodes. The phase arguments of the vector $\ket{x}$ are encoded through z-rotations of angles $\lambda_{j,v}$, and the magnitudes through y-rotations of angles $\alpha_{j,v}=2\asin(\beta_{j,v})$. 

Algorithms \ref{alg:state_decomposition2} and \ref{alg:angle_tree2} describe the construction of a state tree and an angle tree. Respective pseudocodes \ref{alg:state_decomposition} and \ref{alg:angle_tree} are presented in the appendix.

\begin{algorithm}
    Initialize the state tree by the leafs, where each node value is a complex amplitude from a $2^n$ length state vector \\
    Set $k=1$ \\
    \label{st:state_decomposition2} Create a new level with $2^{n-k}$ nodes, where each node $i$ value is $\eta_{i,k}e^{i\Omega_{i,k}}$ (Eq.~\eqref{eq:Omega} and Eq.~\eqref{eq:eta}, $i=1,\dots, 2^{n-k}$) \\
    If $k<n$, set $k=k+1$ and return to Step~\ref{st:state_decomposition2}, otherwise output the state tree

    \caption{State tree construction}
    \label{alg:state_decomposition2}
\end{algorithm}

\begin{algorithm}
    Set $v=n$ \\
    \label{st:angle_tree2} Create a new level with $2^{n-v}$ nodes, where each node $j$ value is $\alpha_{j,v}e^{i\lambda_{j,v}}$ (Eq.~\eqref{eq:beta} and Eq.~\eqref{eq:lambda}, $j=1,\dots, 2^{n-v}$) using data from a state tree generated by Alg.~\ref{alg:state_decomposition2} \\
    If $v>1$, set $v=v-1$ and return to Step~\ref{st:angle_tree2}, otherwise output the angle tree 
    \caption{Angle tree construction}
    \label{alg:angle_tree2}
\end{algorithm}

\subsection{Top-down approach}
\label{sec:top_down}

\begin{figure}[ht]
    \centering
    \resizebox{0.6\columnwidth}{!}
    {
    \begin{quantikz}[column sep=0.15cm]
        \lstick{$\ket{0}_0$}    & \gate[1]{R_y(\alpha_{1,3})} & \octrl{1}                   & \ctrl{1}                    & \octrl{1}
                              & \octrl{1}                   & \ctrl{1}                    & \ctrl{1}                  &\qw &\qw&\qw & \qw\\
        \lstick{$\ket{0}_1$}    & \qw                         & \gate[1]{R_y(\alpha_{1,2})} & \gate[1]{R_y(\alpha_{2,2})} & \octrl{1}
                              & \ctrl{1}                    & \octrl{1}                   & \ctrl{1}                    &
                                \qw &\qw&\qw & \qw\\
        \lstick{$\ket{0}_2$}    & \qw                         & \qw                         & \qw                         & \gate[1]{R_y(\alpha_{1,1})} 
                              & \gate[1]{R_y(\alpha_{2,1})} & \gate[1]{R_y(\alpha_{3,1})} & \gate[1]{R_y(\alpha_{4,1})} &
                                \qw                         & \qw                         & \qw
    \end{quantikz}
    }
    \caption{Quantum circuit to load an 8-dimensional real vector in a quantum device using the top-down amplitude encoding strategy~\cite{shende2006synthesis,mottonen_transformation_2005,bergholm_quantum_2005} (Alg.~\ref{alg:top_down2}). The qubit index indicated by the subscript corresponds to the tree level in Fig.~\ref{fig:angletree1}.}
    \label{fig:mottonen}
\end{figure}
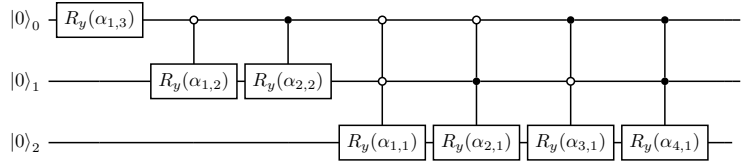

The top-down amplitude encoding approach to quantum state initialization is a linear transformation consisting of a sequence of uniformly controlled rotations~\cite{bergholm_quantum_2005,mottonen_transformation_2005} that takes the initial basis vector $\ket{0}^{\otimes N}$ to some arbitrary vector $\ket{x}=(|x_0|e^{i\omega_0}, \dots, |x_{N-1}|e^{i\omega_{N-1}})^T$. This generates a quantum circuit with complexity of $O_{d}(N)$ and $O_w(\log_2(N))$~\cite{ shende2006synthesis,mottonen_transformation_2005,bergholm_quantum_2005}.

The top-down state preparation (TDSP) algorithm begins by preparing the following state at the root ($v=n$) of the angle tree (see Fig.~\ref{fig:angletree1} for an example)
\begin{equation} \label{eq:mottonenstart}
\ket{\psi_n}=e^{-i\frac{\lambda_{1,n}}{2}}\sqrt{1-|\beta_{1,n}|^2}\ket{0}+e^{i\frac{\lambda_{1,n}}{2}}\beta_{1,n}\ket{1}.
\end{equation}
To load states into the next level (indicated by $v$ in Eq.~\eqref{eq:mottonensum}), the current state (indicated by $v+1$ because $v$ is in reverse order, decreasing from $n$ to $1$) is sequentially combined with the values of the next state in Eq.~\eqref{eq:mottonensum}.
\begin{equation} \label{eq:mottonensum}
\ket{\psi_v}=\left[ \sum_{j=1}^{2^{n-v}} \left( e^{-i\frac{\lambda_{j,v}}{2}}\sqrt{1-|\beta_{j,v}|^2}\ket{0}+e^{i\frac{\lambda_{j,v}}{2}}\beta_{j,v}\ket{1} \right)\ket{j-1}\bra{j-1}\ket{\psi_{v+1}} \right]
\end{equation}
The update of state $\ket{\psi_v}$ is repeated for $v=(n-1), \dots, 1$, thereby obtaining the desired state 
\begin{equation*}
    \ket{\psi_1}=|x_0|e^{i\omega_0}\ket{0}+ \ldots+ |x_{N-1}|e^{i\omega_{N-1}}\ket{N-1}.
\end{equation*}
The summation in Equation~\eqref{eq:mottonensum} expresses the sequential characteristic of the top-down approach, since the state of each layer of the tree needs to be loaded on one qubit through a sequence of rotations. Figure~\ref{fig:mottonen} presents an example quantum circuit for encoding 8-dimensional vector using the top-down state preparation method.

\begin{algorithm}
    Generate a state tree from the input vector \\
    Generate an angle tree from the state tree \\
    Create a quantum circuit with $n$ qubits (one qubit for each angle tree level) \\
    \label{st:top_down1} Perform one y-rotation and one z-rotation on the first qubit (qubits are 0-indexed) using the angle tree root values $\alpha_{1,n}$ and $\lambda_{1,v}$ (Eq.~\eqref{eq:mottonenstart}) \\
    Set $v=n-1$ (starts at the root) \\
    \label{st:top_down2} Perform a \textit{Uniformly Controlled Rotation} controlled by qubits $0,1,\dots,n-v-1$ (corresponding to the previous levels) with the current qubit $n-v$ as target, using the current level nodes values $\alpha_{j,v}$ and $\lambda_{j,v}$ ($1\le j \le 2^{n-v}$) as rotation angles (Eq.~\eqref{eq:mottonensum}) \\
    If $v>1$, set $v=v-1$ and return to Step~\ref{st:top_down2}, otherwise output the encoding quantum circuit
    
    \caption{Top-down state preparation}
    \label{alg:top_down2}
\end{algorithm}

The name \textit{top-down} comes from the way this approach walks through the tree from the root to the leaves to build a quantum circuit. The combination of states is done with multi-controlled rotations, and it takes $\log_2(N)$ qubits to generate the complete state. At each level, it assembles a sequence of rotations targeting one qubit and is controlled by the qubits of the previous levels. First, y-rotations are applied to set the magnitudes, followed by z-rotations to set the phases. These steps are presented in Algorithm \ref{alg:top_down2} with its Pseudocode \ref{alg:top_down} provided in the appendix. 

\subsection{Bottom-up approach}
\label{sec:bottom_up}

\begin{figure}[ht]
    \centering
\begin{subfigure}[b]{.49\textwidth}
    \centering
    \resizebox{1.0\columnwidth}{!}
    {
    \begin{quantikz}
        \lstick{$\ket{0}_0$}    & \gate[1]{R_y(\alpha_{1,3})} \gategroup[7,steps=2,style={dashed,rounded corners,draw=green!60,fill=green!5,inner xsep=3pt}, background,label style={label position=below,yshift=-0.4cm}]{one-qubit states} & \gate[1]{R_z(\lambda_{1,3})} & \qw & \qw    \gategroup[7,steps=4,style={dashed,rounded corners,draw=blue!60,fill=blue!5,inner xsep=3pt}, background,label style={label position=below,yshift=-0.4cm}]{combining states}                    &\qw                         & \ctrl{1}                   & \ctrl{3} &\qw \rstick[wires=2]{output}  \\
        \lstick{$\ket{0}_1$}    & \gate[1]{R_y(\alpha_{1,2})} & \gate[1]{R_z(\lambda_{1,2})} & \qw & \ctrl{2}                   &\qw                         & \swap{1}                   &\qw    &\qw \\
        \lstick{$\ket{0}_2$}    & \gate[1]{R_y(\alpha_{2,2})} & \gate[1]{R_z(\lambda_{2,2})} & \qw &\qw                         & \ctrl{3}                   & \targX{}                   &\qw    &\qw \rstick[wires=1]{ancilla} \\
        \lstick{$\ket{0}_3$}    & \gate[1]{R_y(\alpha_{1,1})} & \gate[1]{R_z(\lambda_{1,1})} & \qw & \swap{1}                   &\qw                         &\qw                         & \swap{2}   &\qw \rstick[wires=1]{output}\\
        \lstick{$\ket{0}_4$}    & \gate[1]{R_y(\alpha_{2,1})} & \gate[1]{R_z(\lambda_{2,1})} & \qw & \targX{}                   &\qw                         &\qw                         &\qw &\qw  \rstick[wires=3]{ancilla}  \\
        \lstick{$\ket{0}_5$}    & \gate[1]{R_y(\alpha_{3,1})} & \gate[1]{R_z(\lambda_{3,1})} & \qw &\qw                         & \swap{1}                   &\qw                         & \targX{}  &\qw  \\
        \lstick{$\ket{0}_6$}    & \gate[1]{R_y(\alpha_{4,1})} & \gate[1]{R_z(\lambda_{4,1})} & \qw &\qw                         & \targX{}                   &\qw                         &\qw &\qw 
    \end{quantikz}
    }
    \caption{}
    \label{fig:dcsp}
\end{subfigure}
\begin{subfigure}[b]{.49\textwidth}
  \resizebox{1.0\columnwidth}{!}
    {
    \begin{quantikz}[row sep=0.2cm]
        \lstick{$a\ket{0}+b\ket{1}$}     & \ctrl{1} & \ctrl{2} & \qw        & \ctrl{4} & \qw \rstick[wires=9]{$a\ket{\psi}\ket{\phi}+b\ket{\phi}\ket{\psi}$} \\
        \lstick[wires=4]{$\ket{\psi}_m$} & \swap{4} & \qw      & \qw        & \qw      & \qw \\
                                         & \qw      & \swap{4} & \qw        & \qw      & \qw \\
                                         &          &          & \vdots     &          & \vdots \\
                                         & \qw      & \qw      & \qw        & \swap{4} & \qw \\
        \lstick[wires=4]{$\ket{\phi}_m$} & \targX{} & \qw      & \qw        & \qw      & \qw \\
                                         & \qw      & \targX{} & \qw        & \qw      & \qw \\
                                         &          &          & \vdots     &          & \vdots \\
                                         & \qw      & \qw      & \qw        & \targX{} & \qw \\
    \end{quantikz}
    }
    \caption{}
   \label{fig:combining}
\end{subfigure}
\caption{Divide-and-conquer bottom-up load strategy. (a) Circuit generated by the divide-and-conquer~\cite{araujo_divide-and-conquer_2021} bottom-up strategy (Alg.~\ref{alg:bottom_up2}) to load an 8-dimensional complex vector in a quantum device. The indexes of the qubits correspond to the tree nodes indexes in Fig.~\ref{fig:angletree1}. The circuit starts with the simultaneous preparation of $(N-1)$ one-qubit states associated with all tree nodes, followed by the combination of states through CSWAPs. (b) Combining states with controlled-swap operations. }
\label{fig:dcsp_strategy}
\end{figure}
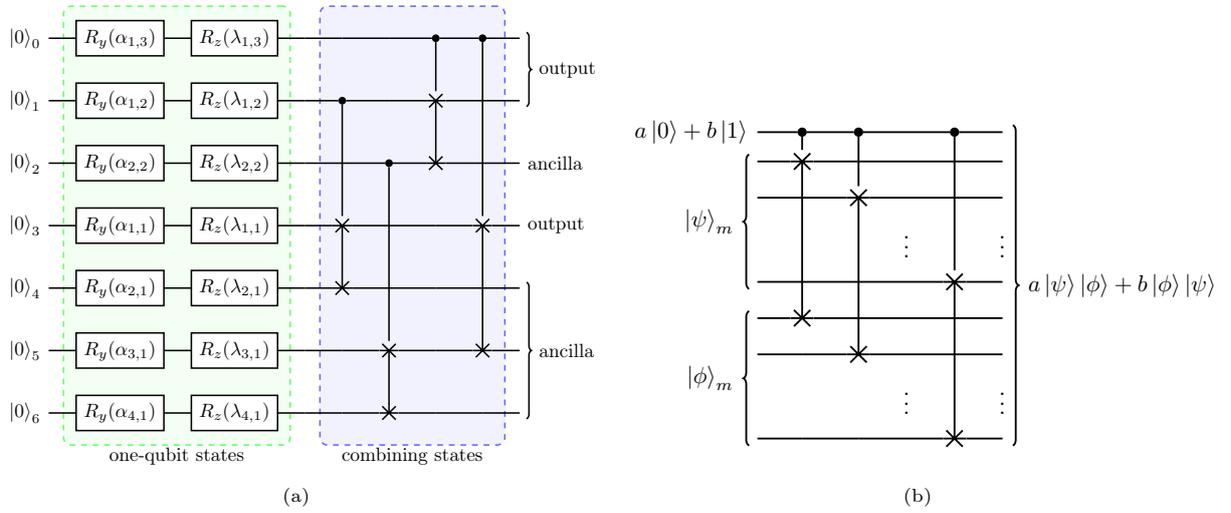

The bottom-up state preparation algorithm constructs a quantum circuit with complexity $O_{d}(\log_2^2(N))$ and $O_w(N)$~\cite{araujo_divide-and-conquer_2021}.
It starts by preparing $N/2$ single-qubit states, corresponding to the leaves of the tree (Fig.\ref{fig:statetree1}). Equations \eqref{eq:lambda} and \eqref{eq:beta} are used starting from the lowest level of the tree ($v=1$), which corresponds to starting from the initial state
\begin{equation} \label{eq:dcleaf_nodes}
    \ket{\psi_{j,1}}=e^{-i\frac{\lambda_{j,1}}{2}}\sqrt{1-|\beta_{j,1}|^2}\ket{0} + e^{i\frac{\lambda_{j,1}}{2}}\beta_{j,1}\ket{1}.
\end{equation}
Loading the states in the upper levels of the tree is done by recursive updates of
\begin{equation}
\begin{split}
    \label{eq:dccombine_children}
    \ket{\psi_{j,v}} =  e^{-i\frac{\lambda_{j,v}}{2}}\sqrt{1-|\beta_{j,v}|^2}\ket{\psi_{2j-1,v-1}}\ket{\psi_{2j,v-1}} +  e^{i\frac{\lambda_{j,v}}{2}}\beta_{j,v}\ket{\psi_{2j,v-1}}\ket{\psi_{2j-1,v-1}},
\end{split}
\end{equation}
where $v=2,\dots,n$. The desired state, with ancilla $\ket{\phi}$, is obtained when $v=n$ as
\begin{equation}
\begin{split}
    \ket{\psi_{1,n}}= |x_0|e^{i\omega_0}\ket{0}\ket{\phi_0}+\dots +  |x_{N-1}|e^{i\omega_{N-1}}\ket{N-1}\ket{\phi_{N-1}}.
\end{split}
\end{equation}

\begin{algorithm}
    Generate a state tree from the input vector \\
    Generate an angle tree from the state tree \\
    Create a quantum circuit with $2^n-1$ qubits (one qubit for each angle tree node) \\
    Perform $2^{n-1}$ y-rotations and z-rotations on qubits $2^{n-1}+j-2$ ($1\le j \le 2^{n-1}$) using the leaf values $\alpha_{j,1}$ and $\lambda_{j,1}$ to prepare $2^{n-1}$ initial single-qubit states (Eq.~\eqref{eq:dcleaf_nodes}, Fig.~\ref{fig:dcsp}) \\
    Set $v=2$ and $j=1$ (starts at the bottom) \\
    \label{st:bottom_up2} Perform one y-rotation and one z-rotation on qubit $2^{n-v}+j-2$ using the node values $\alpha_{j,v}$ and $\lambda_{j,v}$ to prepare a single-qubit state to control \textit{CSWAPs} operations \\
    Perform \textit{Controlled SWAPs} controlled by qubit $2^{n-v}+j-2$ to combine the previous states prepared with the qubits associated to the sub-tree started by the current node (Eq.~\eqref{eq:dccombine_children}, Fig.~\ref{fig:combining}) \\
    If $j<2^{n-v}$, set $j=j+1$ and return to Step~\ref{st:bottom_up2}, otherwise continue \\
    If $v<n$, set $v=v+1$ and return to Step~\ref{st:bottom_up2}, otherwise output the encoding quantum circuit
    
    \caption{Bottom-up state preparation}
    \label{alg:bottom_up2}
\end{algorithm}

Updating the states in Equation \eqref{eq:dccombine_children} requires a method that entangles each of the two states $\ket{\psi_{2j-1,v-1}}$ and $\ket{\phi_{2j,v-1}}$ to orthonormal subspaces $\ket{0}$ and $\ket{1}$, respectively, with designated amplitudes. As demonstrated by Araujo et al.~\cite{araujo_divide-and-conquer_2021}, $m$ controlled-swap (CSWAP) operations can combine two $m$-qubit states in the form of Equation \eqref{eq:dccombine_children} (see Fig.~\ref{fig:combining}) to encode the desired set of amplitudes in the orthonormal subspaces of the first $m+1$ qubits. Since each node of the level is represented by one qubit, multiple loading within a layer can be performed in parallel. Thus, all states in the given layer can be loaded simultaneously. This is an advantage in comparison to the top-down approach which loads each node state sequentially. Since the underlying idea of the bottom-up approach is recursive combination of single-qubit states that are easy to prepare, it was named as divide-and-conquer state preparation (DCSP) when first introduced~\cite{araujo_divide-and-conquer_2021}. An example quantum circuit for encoding $8$-dimensional vector using the DCSP method is depicted in Fig.~\ref{fig:dcsp}. Algorithm \ref{alg:bottom_up2} describes these steps and Pseudocode \ref{alg:bottom_up} is provided in the appendix.

\section{Bidirectional quantum state preparation}
\label{sec:bidirectional}

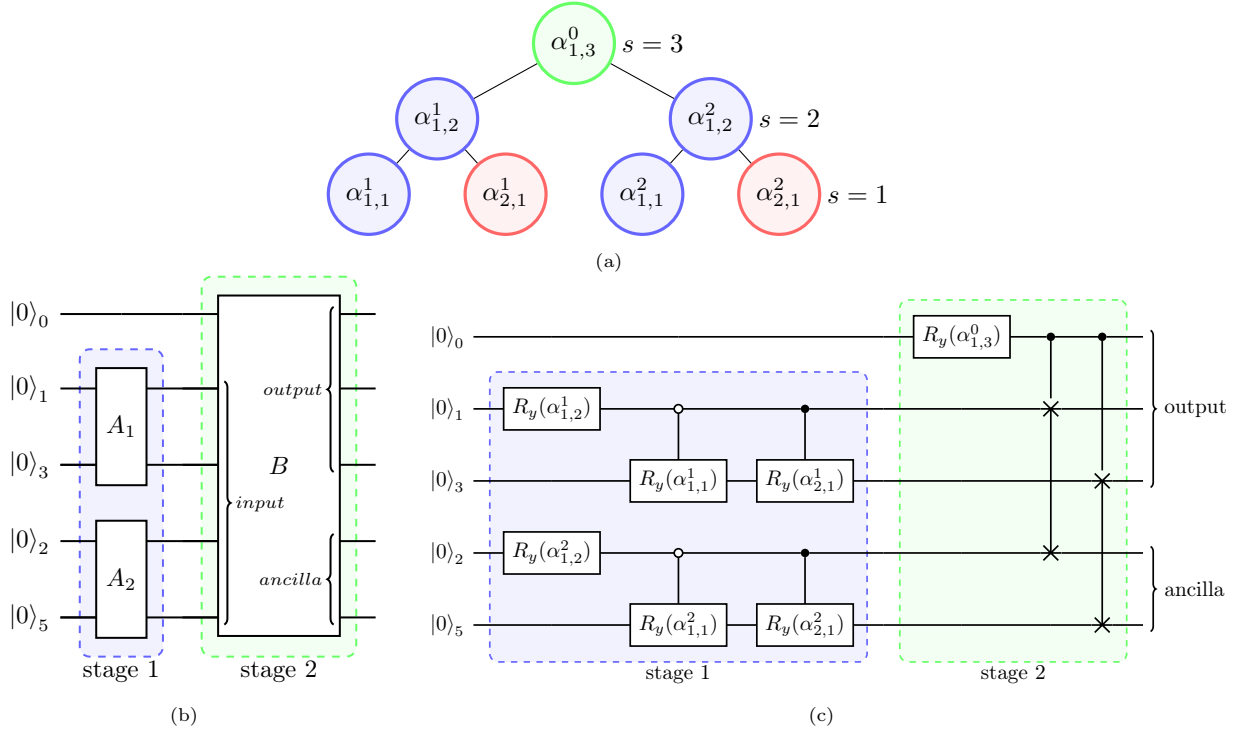
\begin{figure}[ht]
\centering
\begin{subfigure}[b]{.49\textwidth}
  \centering
    \begin{tikzpicture}[level distance=1cm,
        level 1/.style={sibling distance=3.6cm, level distance=1cm},
        level 2/.style={sibling distance=1.8cm, level  distance=1cm},
        level 3/.style={sibling distance=.9cm, level  distance=1.0cm},
        stage2/.style={circle, draw=green!60, fill=green!5, very thick, minimum size=7mm},
        stage1/.style={circle, draw=blue!60, fill=blue!5, very thick, minimum size=7mm},
        remove/.style={circle, draw=red!60, fill=red!5, very thick, minimum size=7mm},
        every label/.append style = {label distance=1pt, inner sep=1pt, align=left}]
    \node[stage2,label={right:$s=3$}]{$\alpha^0_{1,3}$}
        child{
        	node[stage1]{$\alpha^1_{1,2}$}
            child {
                node[stage1] {$\alpha^1_{1,1}$}
        	}
            child {
                node[remove] {$\alpha^1_{2,1}$}
            }
        }
        child{
        node[stage1,label={right:$s=2$}]{$\alpha^2_{1,2}$}
            child {
                node[stage1] {$\alpha^2_{1,1}$}
            }
            child {
                node[remove,label={right:$s=1$}] {$\alpha^2_{2,1}$}
            }
        };
    \end{tikzpicture}
    
    \caption{}
    \label{fig:angletree}
\end{subfigure}

\begin{subfigure}[b]{.32\textwidth}
    \centering
    \resizebox{1.0\columnwidth}{!}
    {
    \begin{quantikz}
        \lstick{$\ket{0}_0$}    & \qw                 & \qw & \gate[wires=5][1.7cm]{B} \gateoutput[wires=3]{$output$} \gategroup[5,steps=1,style={dashed,rounded corners,draw=green!60,fill=green!5,inner xsep=3pt}, background,label style={label position=below,yshift=-0.4cm}]{stage 2} & \qw \\
        \lstick{$\ket{0}_1$}    & \gate[wires=2]{A_1} \gategroup[4,steps=1,style={dashed,rounded corners,draw=blue!60,fill=blue!5,inner xsep=3pt}, background,label style={label position=below,yshift=-0.4cm}]{stage 1} & \qw & \qw \gateinput[wires=4]{$input$}   & \qw \\
        \lstick{$\ket{0}_3$}    & \qw           & \qw & \qw                             & \qw \\
        \lstick{$\ket{0}_2$}    & \gate[wires=2]{A_2} & \qw & \qw \gateoutput[wires=2]{$ancilla$} & \qw \\
        \lstick{$\ket{0}_5$}    & \qw                 & \qw & \qw                & \qw 
    \end{quantikz}
    }
    \caption{}
    \label{fig:bd_block_diagram}
\end{subfigure}
\hfill
\begin{subfigure}[b]{.66\textwidth}
    \centering
    \resizebox{1.0\columnwidth}{!}
    {
    \begin{quantikz}
        \lstick{$\ket{0}_0$}    & \qw                           & \qw                           & \qw & \qw                           & \gate[1]{R_y(\alpha^0_{1,3})} \gategroup[5,steps=3,style={dashed,rounded corners,draw=green!60,fill=green!5,inner xsep=3pt}, background,label style={label position=below,yshift=-0.4cm}]{stage 2}
                                & \ctrl{1}                      & \ctrl{2}                      & \qw  \rstick[wires=3]{output}  \\
        \lstick{$\ket{0}_1$}    & \gate[1]{R_y(\alpha^1_{1,2})} \gategroup[4,steps=3,style={dashed,rounded corners,draw=blue!60,fill=blue!5,inner xsep=3pt}, background,label style={label position=below,yshift=-0.4cm}]{stage 1} & \octrl{1}                     & \ctrl{1} & \qw                     & \qw
                                & \swap{2}                      & \qw                           & \qw \\
        \lstick{$\ket{0}_3$}    & \qw                           & \gate[1]{R_y(\alpha^1_{1,1})} & \gate[1]{R_y(\alpha^1_{2,1})} & \qw & \qw
                                & \qw                           & \swap{2}                      & \qw \\
        \lstick{$\ket{0}_2$}    & \gate[1]{R_y(\alpha^2_{1,2})} & \octrl{1}                     & \ctrl{1} & \qw                     & \qw
                                & \targX{}                      & \qw                           & \qw \rstick[wires=2]{ancilla} \\
        \lstick{$\ket{0}_5$}    & \qw                           & \gate[1]{R_y(\alpha^2_{1,1})} & \gate[1]{R_y(\alpha^2_{2,1})} & \qw & \qw
                                & \qw                           & \targX{}                      & \qw
    \end{quantikz}
    }
    \caption{}
    \label{fig:bddetail}
\end{subfigure}

\caption{Schematics of the bidirectional algorithm. (a) Angle tree example with a split at level $s=2$. The blue and red nodes ($\alpha^1$ and $\alpha^2$) correspond to the bidirectional procedure first stage. In each of the two sub-trees of the first stage, 4 of the 8 amplitudes expected as input by stage 2 are encoded using a top-down method. The green node ($\alpha^0$) above the tree split correspond to the second stage single sub-tree, subject to a partial DCSP bottom-up procedure. The first stage red nodes ($j > 1$) are no longer associated with an ancilla since they are now encoded through a top-down approach.
(b) Block diagram circuit, corresponding to the tree in (a). In stage 1, the $A_k$ operators (the index $k$ is related to angle vectors $\alpha^k$ upper index) are responsible for encoding the amplitudes that will be used as input by stage 2. In this example, each $A_k$ operator encodes 4 amplitudes from a total of 8. The $B$ operator is the partial DCSP for 8 amplitudes, which is initialized with the expected state for the split level 2 and continues with the traditional algorithm.
(c) Detailed view of (b), generated by the bidirectional strategy described in Algorithm \ref{alg:bidirectional2} for a real and positive 8-dimensional input vector. }
\label{fig:bd_data_representation}
\end{figure}

This section presents a bidirectional state preparation (BDSP) method combining both bottom-up and top-down strategies as walking on the tree in both directions. This new strategy can interchange depth and space cost in a configurable manner, thereby allowing for the sublinear cost in both quantum circuit depth and width. In particular, the equilibrium point between these costs achieves the quadratic reduction in both space and time. The algorithm is depicted in Fig.~\ref{fig:bd_data_representation} and the detailed explanation is provided as follows.

The bidirectional state preparation algorithm starts by informing a level $v=s$ (enumerated from bottom to top, where $1\le s \le n$) at which the angle tree is split, followed by two stages. In the first stage, it segments the tree section below $s$ into $2^{n-s}$ sub-trees of height $s$. The $2^{n-s}$ nodes at level $s$ are the roots of these sub-trees. The number of sub-trees determines how many initial sub-states should be prepared in the first stage of the algorithm. The amplitude values of these sub-states $a_j=(a_{j,1}, \dots, a_{j, 2^s})$ ($1 \le j \le 2^{n-s}$) are loaded concurrently using a sequential algorithm~\cite{shende2006synthesis,mottonen_transformation_2005,bergholm_quantum_2005} based on the TDSP method as
\begin{equation}
    \label{eq:ampencoded}
    \ket{\psi_{j,s}} = \sum_{k=1}^{2^s} a_{j,k}\ket{k-1} \quad  ; \quad j=1, 2,\dots, 2^{n-s}.
\end{equation}
The initial sub-states are the input of the second stage of BDSP. They reproduce the state that would be created by the bottom-up steps up to the split level $s$. In the second stage, the sub-states are combined to generate the complete state by the divide-and-conquer approach (Fig.~\ref{fig:bddetail}). The bottom-up algorithm takes the state prepared in the first stage as the input, and starts walking on the tree from the split level (Eq.~\eqref{eq:dccombine_children}, where $v=s+1, \dots, n$). In other words, the BDSP follows the bottom-up DCSP algorithm starting from states $\ket{\psi_{j,s}}$ (Eq.~\eqref{eq:ampencoded}) instead of starting from the single-qubit leaf states (Eq.~\eqref{eq:dcleaf_nodes}). 
The BDSP algorithm is described in Algorithm \ref{alg:bidirectional2} below with Pseudocode \ref{alg:bidirectional} provided in the appendix.

\begin{algorithm}
    Generate a state tree from the input vector \\
    Generate an angle tree from the state tree \\
    Create a quantum circuit with $(s+1)2^{n-s}-1$ qubits (Eq.~\eqref{eq:space}) \\
    Perform Algorithm \ref{alg:top_down2} (top-down approach) starting from step \ref{st:top_down1} to prepare $2^{n-s}$ states of $s$-qubits (replacing $n$ by $s$), using the $2^{n-s}$ sub-trees as input for each state (Eq.~\eqref{eq:ampencoded}). This step is named \textit{Stage 1} \\
    Perform Algorithm \ref{alg:bottom_up2} (bottom-up approach) starting from step \ref{st:bottom_up2} and $v=s+1$ to combine the $2^{n-s}$ states prepared in \textit{Stage 1} using the remaining $2^{n-s}-1$ qubits (Eq.~\eqref{eq:dccombine_children}). This step is named \textit{Stage 2} \\
    Output the encoding quantum circuit
    
    \caption{Bidirectional state preparation}
    \label{alg:bidirectional2}
\end{algorithm}

\subsection{Complexity}
\label{sec:complexity}
In general, the BDSP algorithm builds quantum circuits whose depth and width are expressed respectively by
\begin{equation} \label{eq:depth}
    \underbrace{\frac{N}{2^{\log_2(N)-s}} }_\text{stage 1} + \underbrace{\sum_{i=s+1}^{\log_2(N)} i-1 }_\text{stage 2} 
    = 2^{s} +\frac{1}{2} (\log_2^2(N)- \log_2(N) - s^2 + s)  
\end{equation}
and
\begin{equation} \label{eq:space}
    \underbrace{s\frac{N}{2^{s}}}_\text{stage 1} + 
    \underbrace{\frac{N}{2^{s}} - 1}_\text{stage 2} = (s+1)\frac{N}{2^{s}} - 1,
\end{equation}
where $N$ is the number of amplitudes (i.e. the dimension of the data vector) and $s$ is a parameter indicating the tree splitting level (the tree level in reverse order).
Stage 1 and 2 indicate the contribution from each stage of the bidirectional procedure to the circuit complexity stated in Theorem~\ref{theo:depth_width}.

\begin{theorem} \label{theo:depth_width}
Algorithm~\ref{alg:bidirectional2} generates a quantum circuit with depth $O_d \left(2^{s} + \log_2^2 \left( N \right)-s^2 \right)$ and width \\ $O_w\left( \left(s+1\right)\frac{N}{2^{s}}\right)$.
\end{theorem}
In Equation \eqref{eq:depth}, first term (stage 1) is the leading-order approximation of the quantum circuit depth from existing top-down based algorithms~\cite{mottonen_transformation_2005,shende2006synthesis} for sub-states with $s$ qubits. The exact expression depends on which of the two algorithms is used.
The summation of the second term (stage 2) is the divide-and-conquer circuit depth from split level $s+1$ to $n$.
Similarly, the first term in Equation \eqref{eq:space} is the number of qubits occupied by all first stage sub-states and
the second term is the number of qubits used by the second stage.

There are three noteworthy configuration values for the parameter $s$ (see Table~\ref{tab:complexity}). Setting $s=\log_2(\sqrt{N})$ achieves asymptotic sublinearity, and $s=1$ or $s=\log_2(N)$ recovers bottom-up or top-down approaches.

\begin{table}[ht]
	\centering
	\begin{tabular}{cccc}
		\hline
		& \begin{tabular}{@{}c@{}}bottom-up \\ $s=1$\end{tabular} & \begin{tabular}{@{}c@{}}top-down \\ $s=n$\end{tabular} & \begin{tabular}{@{}c@{}}sublinear \\ $s=n/2$\end{tabular}  \\
    	\hline
		$O_d$ & $n^2$ & $2^n$ & $2^{n/2}$ \\
		$O_w$ & $2^n$ & $n$ & $2^{n/2}$ \\
		\hline
	\end{tabular}
	\caption{Bidirectional quantum circuit complexity for different configurations. These expressions were obtained from Eq.~\eqref{eq:depth} and Eq.~\eqref{eq:space}.}
	\label{tab:complexity}
\end{table}

The condition for quadratic reduction in both depth and width is obtained through asymptotic analysis of the minimum distance between Eq.~\eqref{eq:depth} and Eq.~\eqref{eq:space}. The first (second) equation is a monotonically increasing (decreasing) function $\forall s \in \{x \in \mathbb{R} | 4 \le x \le \log_2(N)\}$ and there is a point $s$ where the distance is zero when $N\rightarrow \infty$. Thus the minimum distance point is given by finding $s$ that satisfies
\begin{equation}  \label{eq:asymptotic}
    \lim_{N \rightarrow \infty} O_w-O_{d}=0.
\end{equation}  

\begin{figure}[ht]
    \centering
    \includegraphics[width=0.5\textwidth]{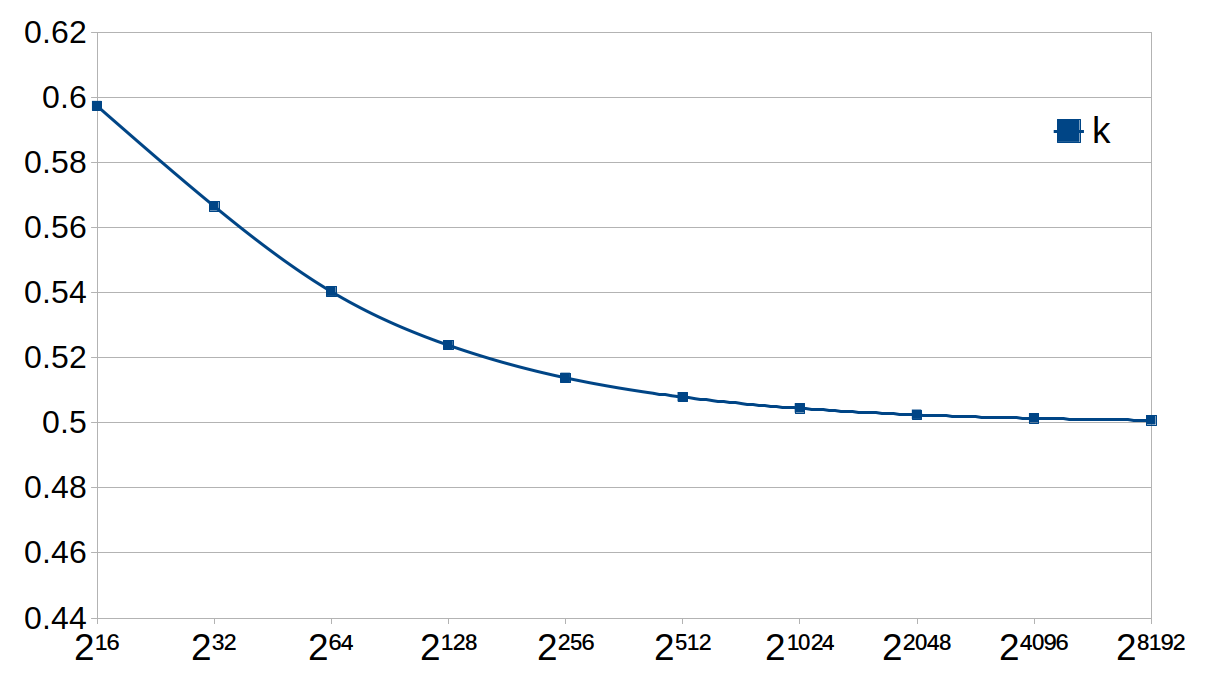}
    \caption{The solution of the system $O_w-O_d=0$ approaches $k=\nicefrac{1}{2}$ as $N$ increases.}
	\label{fig:asymptotic_approximation}
\end{figure}

The asymptotic analysis starts by rewriting Eq.~\eqref{eq:depth} and Eq.~\eqref{eq:space} using a more convenient parameterization,
\begin{equation*}
    s=f(k)=k\log_2(N) \quad \text{where} \quad k \in \left[\frac{4}{\log_2(N)}, 1\right].
\end{equation*}
Applying the limit of Eq.~\eqref{eq:asymptotic} results in the following simplified expression
\begin{align*}
    N^{2k-1} = 1.
\end{align*}
Solving the above equation for $k$ gives the solution $k=1/2$. Therefore, to achieve sublinear circuit complexity with quadratic reductions in both quantum circuit depth and width, the tree split must occur at $s=\nicefrac{1}{2}\log_2(N)=\nicefrac{n}{2}$, which leads to Theorem \ref{theo:sublinearity}.
\begin{theorem} \label{theo:sublinearity}
Algorithm~\ref{alg:bidirectional2} with $s=\nicefrac{n}{2}$ and $N\gg 1$ generates a quantum circuit with sublinear depth $O_{d}\left( \sqrt{N} \right)$ and width $O_w\left( \sqrt{N} \right)$.
\end{theorem}
When dealing with input vectors of small size, $s$ can be calculated by solving Eq.~\eqref{eq:asymptotic} directly with $N$ being a constant. If $s$ cannot be calculated exactly, it can be approximated with the asymptotic result $s=\lceil \nicefrac{n}{2} \rceil$. The reason for the \textit{ceiling} function is because $s$ approximates $\nicefrac{n}{2}$ from upper values (Fig.~\ref{fig:asymptotic_approximation}).

\begin{corollary}
\label{cor:1}
When $N \le 8$ a top-down approach ($s=\log_2(N)$) should always be used, since space and depth both decrease as $s$ increases in the interval $s \in [1..3]$ (see Table \ref{tab:tradeoff} for a numerical example). Circuit depth increases only when $s \ge 4$.
\end{corollary}

\subsection{Experiment}
\label{sec:experiment}

To evaluate the bidirectional method, proof-of-principle experiments were performed on a classical simulator provided by IBM, four superconducting-qubit based quantum devices provided by IBM, and an ion-trap based quantum device provided by IonQ. These are named as ibmq\_qasm\_simulator, ibmq\_rome, ibmq\_santiago, ibmq\_casablanca, and ibmq\_jakarta, and IonQ, respectively. The experiments aim to load the following 8 and 16-dimensional real input vectors:

\begin{smaller}
\begin{equation*}
    \left(\sqrt{0.03}, \sqrt{0.06}, \sqrt{0.15}, \sqrt{0.05}, \sqrt{0.1}, \sqrt{0.3},\sqrt{0.2}, \sqrt{0.11} \right)
\end{equation*}
\end{smaller}
and
\begin{smaller}
\begin{equation*}
    \left( \sqrt{0.01}, \sqrt{0.02}, \sqrt{0.04}, \sqrt{0.02}, \sqrt{0.07}, \sqrt{0.08}, \sqrt{0.04}, \sqrt{0.01}, \sqrt{0.08}, \sqrt{0.02}, \sqrt{0.21}, \sqrt{0.09}, \sqrt{0.12}, \sqrt{0.08}, \sqrt{0.05}, \sqrt{0.06} \right) .
\end{equation*}
\end{smaller}

\begin{table}[t]
    \resizebox{0.45\textwidth}{!}{
    \begin{tabular}{ccccc}
		\hline
		device & $N$ & $s$ & runs & MAE \\

		\hline
		\multirow{5}{*}{\begin{tabular}{@{}c@{}}ibmq\_qasm\_simulator \\ 32 qubits\end{tabular}} & \multirow{3}{*}{8} & 1 & 5 & 0.0016 \\
		&& 2 & 5 & 0.0005  \\
		&& 3 & 5 & 0.0015  \\
		\cline{2-5}
		& \multirow{2}{*}{16} & 2 & 5 & 0.0010  \\
		&& 4 & 5 & 0.0010  \\
		
		\hline
		\multirow{3}{*}{\begin{tabular}{@{}c@{}}ibmq\_rome \\ 5 qubits\end{tabular}} & \multirow{2}{*}{8} & 2 & 10 & 0.0577  \\
		&& 3 & 10 & \textbf{0.0429} \\
		\cline{2-5}
		& \multirow{1}{*}{16} & 4 & 5 & 0.0409  \\
		
		\hline
		\multirow{3}{*}{\begin{tabular}{@{}c@{}}ibmq\_santiago \\ 5 qubits\end{tabular}} & \multirow{2}{*}{8} & 2 & 10 & 0.0464 \\
		&& 3 & 10 & \textbf{0.0233} \\
		\cline{2-5}
		& \multirow{1}{*}{16} & 4 & 10 & 0.0225  \\
		
		\hline

		\multirow{2}{*}{\begin{tabular}{@{}c@{}}ibmq\_casablanca \\ 7 qubits\end{tabular}} & \multirow{3}{*}{8} & 1 & 10 & 0.0710 \\
		&& 2 & 10 & 0.0691 \\
		&& 3 & 10 & \textbf{0.0213} \\
		
		\hline
		\multirow{2}{*}{\begin{tabular}{@{}c@{}}ibmq\_jakarta \\ 7 qubits\end{tabular}} & \multirow{3}{*}{8} & 1 & 10 & 0.0594 \\
		&& 2 & 10 & 0.0497 \\
		&& 3 & 10 & \textbf{0.0289} \\
		
		\hline
		\multirow{5}{*}{\begin{tabular}{@{}c@{}}IonQ \\ 11 qubits\end{tabular}} & \multirow{3}{*}{8} & 1 & 5 & 0.0455 \\
		&& 2 & 5 & 0.0242 \\
		&& 3 & 5 & \textbf{0.0217} \\
		
		\cline{2-5}
		& \multirow{2}{*}{16} & 2 & 5 & 0.0261 \\
		&& 4 & 5 & \textbf{0.0107} \\
		
		\hline
	\end{tabular}
	}
    \captionof{table}{Results of the BDSP experiments that encode $N$-dimensional input vectors in the amplitudes of quantum states using a classical simulator and quantum devices for $N=\lbrace 8, 16\rbrace$. The acronym MAE stands for mean absolute error. The bold font indicates the smallest MAE, and hence the best performance, among different configurations of $s$ for each device and input vector.} \label{tab:experiments}
\end{table}

\begin{table}[t]
    \resizebox{\textwidth}{!}{
    \begin{tabular}{c|ccc|ccc|ccc|ccc}\hline
           &       & $N=8$ &        &       & $N=16$ &        &       & $N=32$&        &       & $N=64$&       \\
        s  & CNOTs & depth & qubits & CNOTs & depth  & qubits & CNOTs & depth & qubits & CNOTs & depth & qubits \\ \hline
        1  & 28    & 31    & 7      & 77    & 58     & 15     & 182   & 93    & 31     & 399   & 136   & 63     \\
        2  & 18    & 24    & 5      & 57    & 51     & 11     & 142   & 86    & 23     & 319   & 129   & 47     \\
        3  & 10    & 20    & 3      & 41    & 48     & 7      & 110   & 83    & 15     & 255   & 126   & 31     \\
        4  &       &       &        & 26    & 51     & 4      & 80    & 87    & 9      & 195   & 130   & 19     \\
        5  &       &       &        &       &        &        & 58    & 114   & 5      & 151   & 158   & 11     \\
        6  &       &       &        &       &        &        &       &       &        & 122   & 241   & 6      \\ \hline
    \end{tabular}
    }
    \captionof{table}{Exchange between circuit depth, width (qubits), and number of CNOTs by adjusting the parameter $s$ (split). $s$ can be interpreted as a hyperparameter to fine-tune the encoding circuit to hardware characteristics such as relaxation time, dephasing time, and the CNOT gate error.}
    \label{tab:tradeoff}
\end{table}

\begin{figure}[ht]
    \centering
    \begin{subfigure}[b]{0.237\textwidth}
        \centering
        \includegraphics[width=1.0\textwidth]{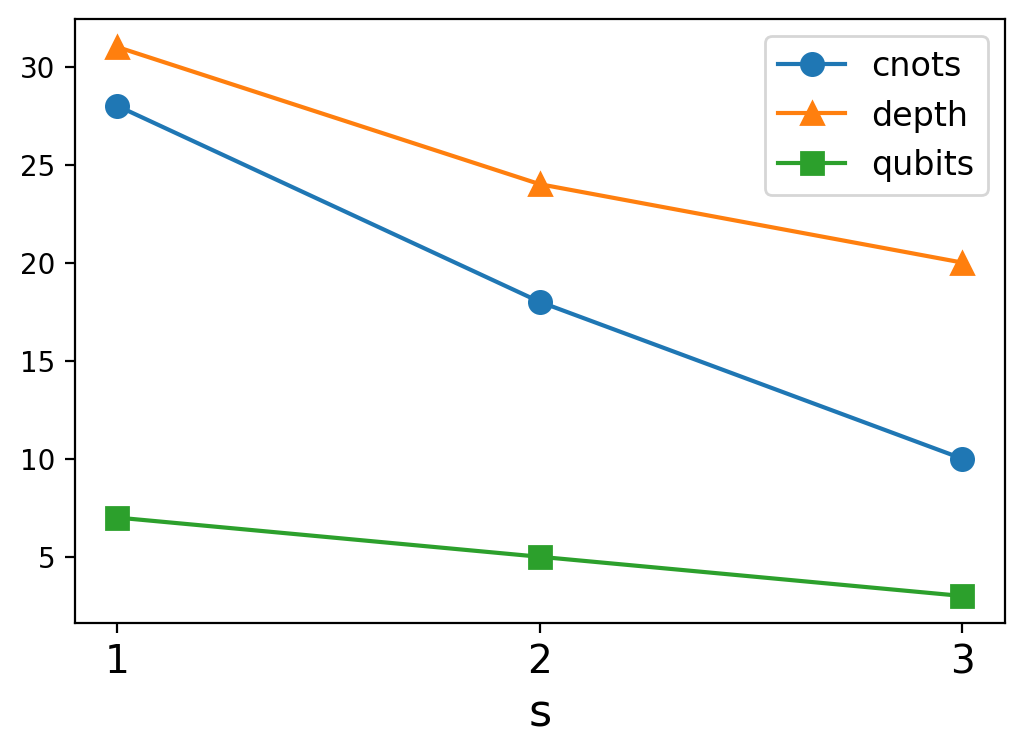}
        \caption{$2^{3}$ complex amplitudes.}
        \label{fig:tradeoff_3}
    \end{subfigure}
    \begin{subfigure}[b]{0.24\textwidth}
        \centering
        \includegraphics[width=1.0\textwidth]{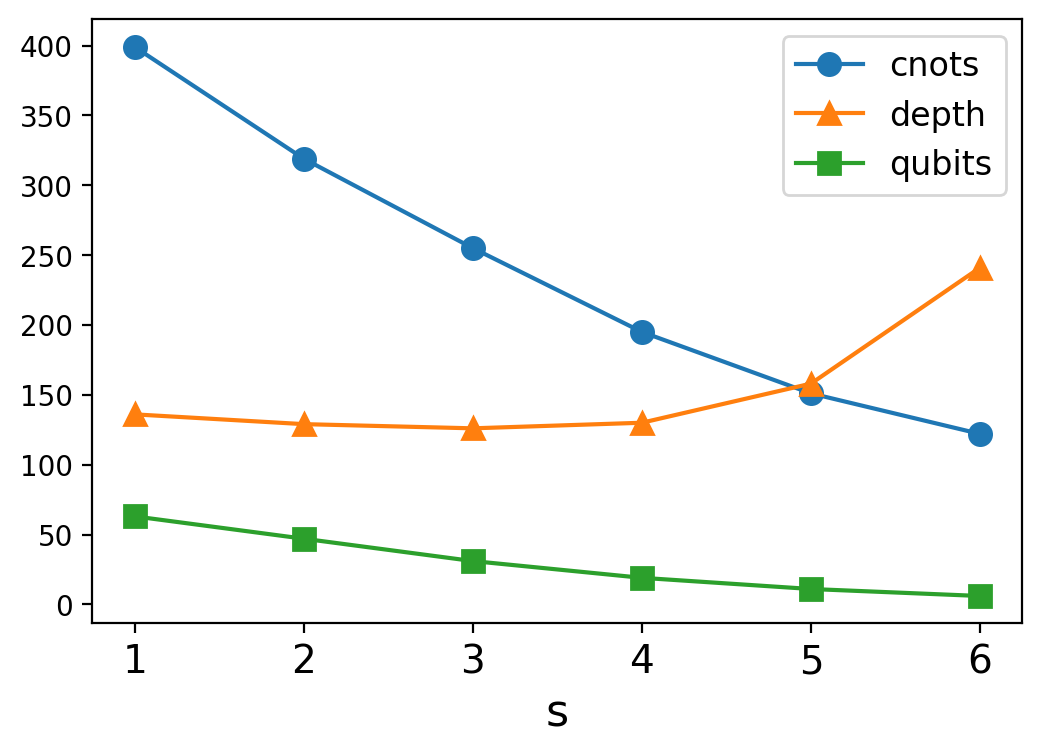}
        \caption{$2^{6}$ complex amplitudes.}
        \label{fig:tradeoff_6}
    \end{subfigure}
    \begin{subfigure}[b]{0.244\textwidth}
        \centering
        \includegraphics[width=1.0\textwidth]{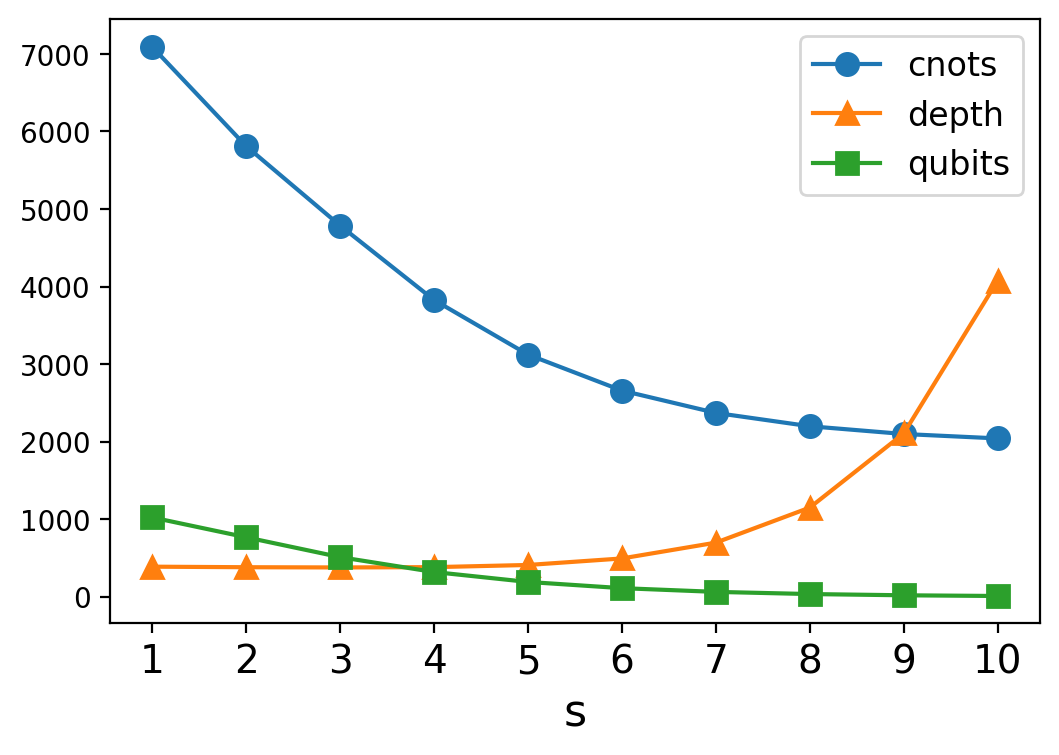}
        \caption{$2^{10}$ complex amplitudes.}
        \label{fig:tradeoff_10}
    \end{subfigure}
    \begin{subfigure}[b]{0.2534\textwidth}
        \centering
        \includegraphics[width=1.0\textwidth]{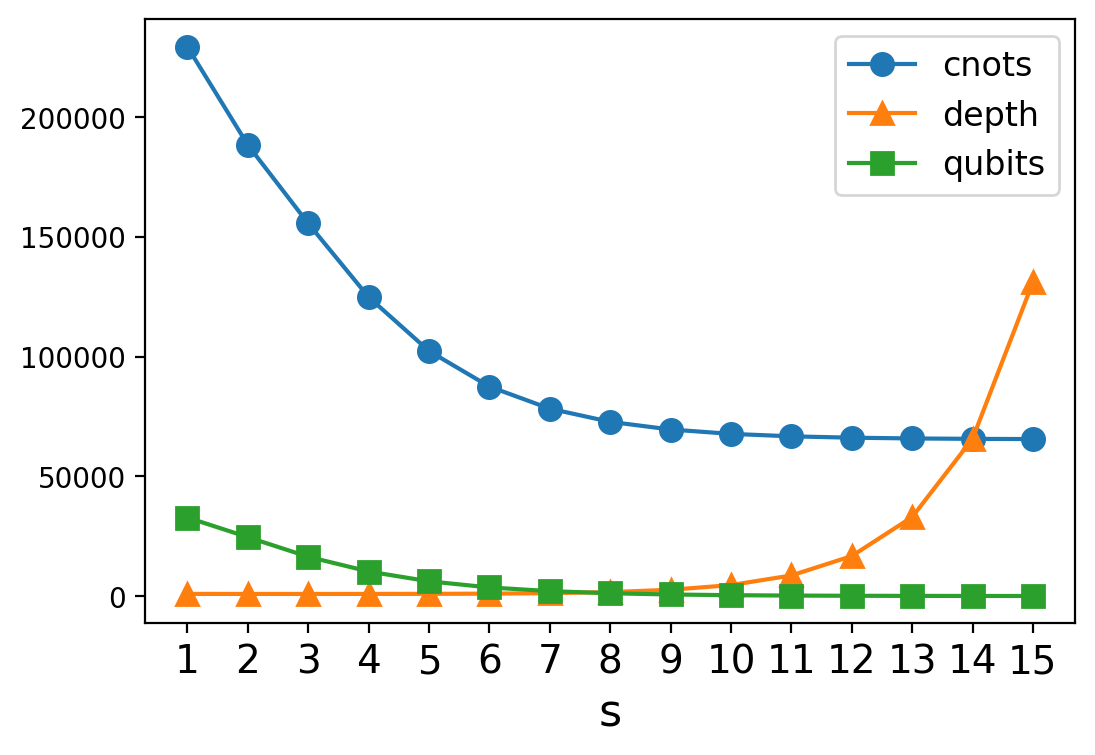}
        \caption{$2^{15}$ complex amplitudes.}
        \label{fig:tradeoff_15}
    \end{subfigure}
    \caption{Exchange between circuit depth, width (number of qubits), and number of CNOTs to load a $2^{n}$-dimensional complex vector into a quantum computer by adjusting parameter $s$. The increasing number of CNOTs at lower depths is a consequence of exchanging computational time for space, given the combination of distant states. It also indicates an increase in concurrent operations. 
    }
    \label{fig:tradeoff}

\end{figure}

\begin{figure}[ht]
\centering
    \begin{subfigure}[b]{.32\textwidth}
        \centering
        \includegraphics[width=1.0\textwidth]{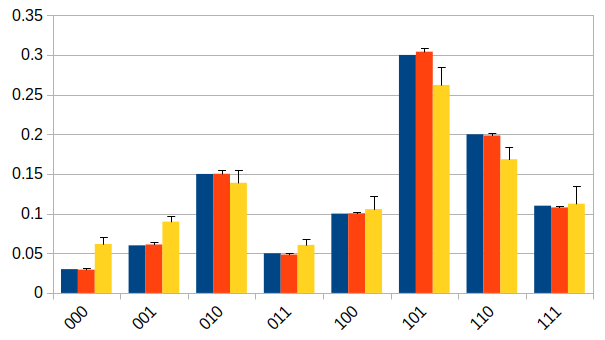}
        \caption{$N=8$ top-down.}
        \label{fig:topdown8}
    \end{subfigure}
    \begin{subfigure}[b]{.32\textwidth}
        \centering
        \includegraphics[width=1.0\textwidth]{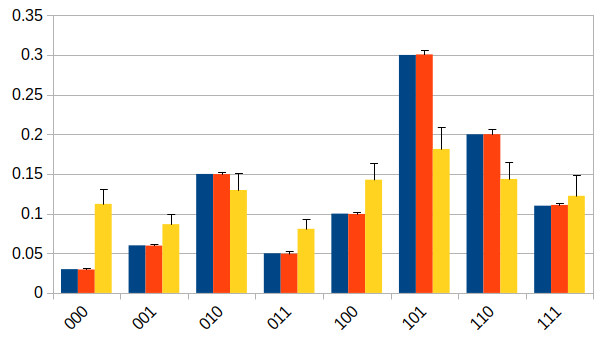}
        \caption{$N=8$ sublinear.}
        \label{fig:sublinear8}
    \end{subfigure}
    \begin{subfigure}[b]{.32\textwidth}
        \centering
        \includegraphics[width=1.0\textwidth]{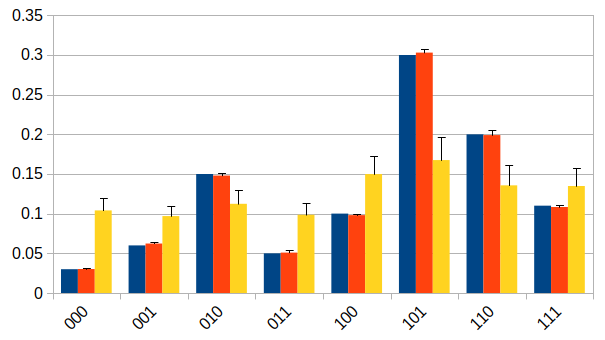}
        \caption{$N=8$ bottom-up.}
        \label{fig:bottomup8}
    \end{subfigure}
    
    \begin{subfigure}[b]{.32\textwidth}
        \centering
        \includegraphics[width=1.0\textwidth]{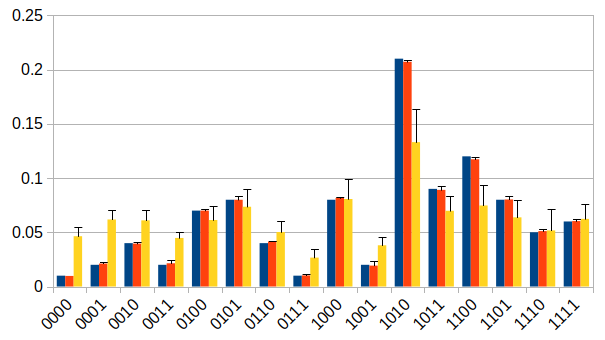}
        \caption{$N=16$ top-down.}
        \label{fig:topdown16}
    \end{subfigure}
    \begin{subfigure}[b]{.32\textwidth}
        \centering
        \includegraphics[width=1.0\textwidth]{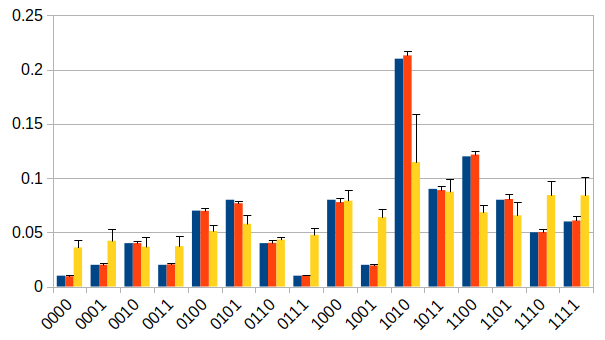}
        \caption{$N=16$ sublinear.}
        \label{fig:sublinear16}
    \end{subfigure}
    \begin{subfigure}[b]{.32\textwidth}
        \centering
        \includegraphics[width=0.5\textwidth]{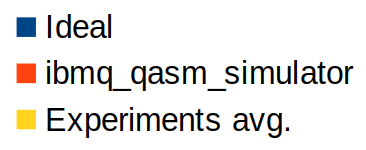} \newline \newline \newline \newline \newline
    \end{subfigure}
    
    \caption{Experimental results with 8- and 16-dimensional input vectors. Blue and red bars indicate respectively the ideal results and the ibm\_qasm\_simulator results. Yellow bars indicate the output average values from the experiments on all quantum devices. Error bars are the standard deviation. 
    }
    \label{fig:experiments}
\end{figure}

Three configurations of the bidirectional method are compared, namely top-down ($s=n$), bottom-up ($s=1$), and sublinear ($s=\lceil \nicefrac{n}{2} \rceil $). The first case uses the least number of qubits $O_w(\log_2(N))$ and maximum depth $O_d(N)$. In the second configuration, depth is minimum $O_d(\log_2^2(N))$ and the number of qubits is maximum $O_w(N)$. The last configuration uses the best trade-off between the quantum circuit depth and width and achieves the sublinear scaling for both. In this case, the quantum circuit depth and width both grow as $O(\sqrt{N})$.

Table \ref{tab:experiments} lists the experimental results, presenting the number of runs per device and dimensionality of the input vector. The ibmq\_rome and ibmq\_santiago devices have only five qubits, and due to this limitation they are not suitable to encode the 8-dimensional vector with the bottom-up configuration (i.e. $s=1$) or to perform sublinear (i.e. $s=\lceil \nicefrac{n}{2} \rceil$) and bottom-up experiments to encode the 16-dimensional vector (see Corollary~\ref{cor:1}). None of the quantum devices used in this work has the capacity to run the bottom-up configuration for the 16-dimensional input vector, which requires at least 15 qubits (i.e. $N-1$ qubits).

Figure \ref{fig:experiments} presents the average output of the experiments with 8 and 16-dimensional input vectors. The height of blue and red bars is an average value obtained from a number of repetitions shown in the \textit{runs} column in Table~\ref{tab:experiments}, and the error bars represent the standard deviation. The height of the yellow bar is the experimental result averaged over all quantum devices.

Table~\ref{tab:tradeoff} and Figure~\ref{fig:tradeoff} show the trade-off between quantum circuit depth, width and the number of CNOT gates as $s$ is varied for randomly generated target vectors of various sizes. 
As expected through the analysis of the number of CNOT gates and the circuit depth in Tab.~\ref{tab:tradeoff}, 
the experimental results in Table~\ref{tab:experiments} and Figure~\ref{fig:experiments} show performance favoring the top-down configuration ($s=n$) for small input sizes ($N<64$) due to the smaller number of CNOT gates and the smaller or approximately equal depth of the circuit. The number of CNOT gates, circuit depth, and number of qubits all decrease as $s$ progresses to $s=3$. The depth starts to increase when $s>3$, as previously implied by Eq.~\eqref{eq:depth}. The comparison employs the mean absolute error (MAE). For each device and input size, the ranking is established where a smaller MAE indicates better performance (see Table~\ref{tab:experiments}).

Data from Table~\ref{tab:tradeoff} and Figure~\ref{fig:tradeoff} were obtained using the \textit{transpile} method in the Quantum Information Science Kit (Qiskit ~\cite{gadi_aleksandrowicz_2019_2562111}) version 0.26.2 to decompose the circuits into physical single-qubit gates and the CNOT gate. These circuits were generated by the bidirectional algorithm with random complex input vectors. The Python code used in this work for implementing Algorithm \ref{alg:top_down} employs functions \textit{ucry} and \textit{ucrz} from Qiskit. These functions are called uniformly controlled rotations (or multiplexers), and the corresponding code in Qiskit is based on the work of Shende et al.~\cite{shende2006synthesis}.

Note that algorithms \ref{alg:top_down2} and \ref{alg:bottom_up2} allocate logical qubits as they are needed without concerning their assignment to physical qubits of the quantum device. For NISQ devices with limited quantum device coupling map, the logical to physical qubit mapping should be optimized in order to minimize the overhead in the quantum circuit depth and the number of gates.

\section{Sparse bidirectional quantum state preparation}
\label{sparse}

The previous section explained the bidirectional method for encoding data in the amplitudes of a quantum state consisting of $n$ qubits, where $2^n=N$ is the total amount of amplitudes of that state. Like other approaches for loading data in the amplitudes, the method presented in this paper has a classical preprocessing complexity $O(N)$. Moreover, such algorithms generate circuits whose spatial and depth costs also depend on $N$. This dependence on the total number of state amplitudes makes these methods inappropriate for encoding sparse input vectors, where the number of non-zero amplitudes $M$ is much smaller than $2^n$.

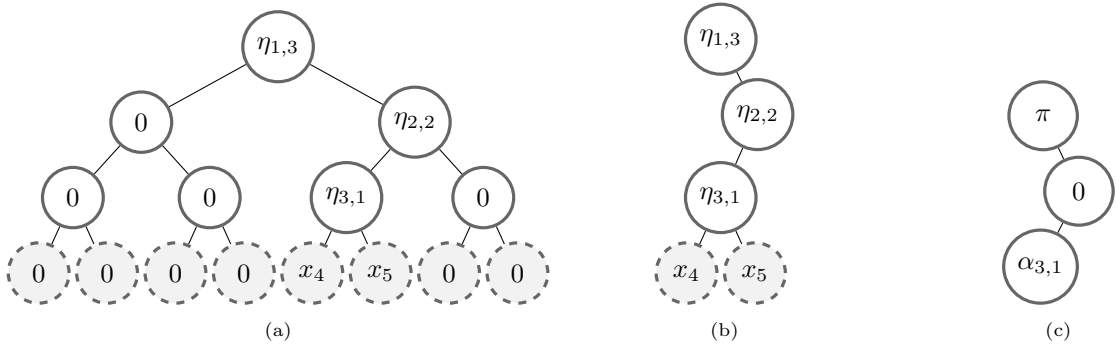
\begin{figure}[t]
    \centering
    \begin{subfigure}[b]{.44\textwidth}
      \centering
        \begin{tikzpicture}[level distance=1cm,
            level 1/.style={sibling distance=3.6cm, level distance=1cm},
            level 2/.style={sibling distance=1.8cm, level  distance=1cm},
            level 3/.style={sibling distance=.9cm, level  distance=1.0cm},
            level 4/.style={sibling distance=.9cm, level  distance=1.0cm},
            circle1/.style={circle, draw=black!60, fill=white!5, very thick, minimum size=8mm},
            circle2/.style={circle, draw=black!60,dashed, fill=black!5, very thick, minimum size=8mm}]
            \node[circle1]{$\eta_{1,3}$}
                child{
                	node[circle1]{$0$}
                    child {
                        node[circle1] {$0$}
                        child {
                            node[circle2] {$0$}
                    	}
                        child {
                            node[circle2] {$0$}
                        }
                	}
                    child {
                        node[circle1] {$0$}
                        child {
                            node[circle2] {$0$}
                    	}
                        child {
                            node[circle2] {$0$}
                        }
                    }
                }
                child{
                node[circle1]{$\eta_{2,2}$}
                    child {
                        node[circle1] {$\eta_{3,1}$}
                        child {
                            node[circle2] {$x_4$}
                    	}
                        child {
                            node[circle2] {$x_5$}
                        }
                    }
                    child {
                        node[circle1] {$0$}
                        child {
                            node[circle2] {$0$}
                    	}
                        child {
                            node[circle2] {$0$}
                        }
                    }
                };
        \end{tikzpicture}
        \caption{}
        \label{fig:statetree2}
    \end{subfigure}
    \begin{subfigure}[b]{.26\textwidth}
      \centering
        \begin{tikzpicture}[level distance=1cm,
            level 1/.style={sibling distance=3.6cm, level distance=1.0cm},
            level 2/.style={sibling distance=1.8cm, level  distance=1.1cm},
            level 3/.style={sibling distance=.9cm, level  distance=1.0cm},
            level 4/.style={sibling distance=.9cm, level  distance=1.0cm},
            circle1/.style={circle, draw=black!60, fill=white!5, very thick, minimum size=8mm},
            circle2/.style={circle, draw=black!60,dashed, fill=black!5, very thick, minimum size=8mm}]
            \node[circle1]{$\eta_{1,3}$}
                child{
                node[circle1,right]{$\eta_{2,2}$}
                    child {
                        node[circle1,left] {$\eta_{3,1}$}
                        child {
                            node[circle2] {$x_4$}
                    	}
                        child {
                            node[circle2] {$x_5$}
                        }
                    }
                };
        \end{tikzpicture}
        \caption{}
        \label{fig:statetree3}
    \end{subfigure}
    \begin{subfigure}[b]{.26\textwidth}
      \centering
        \begin{tikzpicture}[level distance=1cm,
            level 1/.style={sibling distance=3.6cm, level distance=1cm},
            level 2/.style={sibling distance=1.8cm, level  distance=1cm},
            level 3/.style={sibling distance=.9cm, level  distance=1.0cm},
            circle1/.style={circle, draw=black!60, fill=white!5, very thick, minimum size=9mm}]
            \node[circle1]{$\pi$}
                child{
                node[circle1,right]{$0$}
                    child {
                        node[circle1,left] {$\alpha_{3,1}$}
                    }
                };
        \end{tikzpicture}
        \caption{}
        \label{fig:angletree2}
    \end{subfigure}
    \caption{Sparse state preparation. (a) Dense state decomposition of an 8-dimensional input vector with only data patterns $4$ and $5$ not equal to zero. Compare with Figure~\ref{fig:statetree1}. (b) Sparse state decomposition generated by Pseudocode~\ref{alg:sparse_state_decomposition}. The information in this sparse decomposition is equal to that of Figure~\ref{fig:statetree2}. (c) Sparse angle tree generated by function angle\_tree (Pseudocode~\ref{alg:angle_tree}) from the sparse state tree shown in Figure~\ref{fig:statetree3}. Node $\alpha_{1,3}$ is always equal to $\pi$ because $\eta_{1,3}$ only has a child on the right (Pseudocode~\ref{alg:sparse_state_decomposition}, Line~\ref{line:ampright}, and Pseudocode~\ref{alg:angle_tree}, Line~\ref{line:angletreeamp}). Node $\alpha_{2,2}$ is always zero because $\eta_{2,2}$ only has a child on the left (Pseudocode~\ref{alg:sparse_state_decomposition}, Line~\ref{line:ampleft}, and Pseudocode~\ref{alg:angle_tree}, Line~\ref{line:leftchild}). Node $\eta_{3,1}$ has left and right children, so $\alpha_{3,1}$ value depends on nodes $x_4$ and $x_5$ (Pseudocode~\ref{alg:sparse_state_decomposition}, Line~\ref{line:amprightleft}).}
    \label{fig:sparsestatepreparation}
\end{figure}

This section presents a sparse variant of the bidirectional algorithm (Pseudocode~\ref{alg:sparse_bidirectional}). This variant reduces the classical preprocessing complexity to $O(M)$ through a modification in the construction of the state tree (Pseudocode~\ref{alg:sparse_state_decomposition}, Fig.~\ref{fig:statetree3}). Knowing that such a tree is binary, one can see that the value of each node is a combination of the pair at the lower level (see Eq.~\ref{eq:beta}). If the zero-valued amplitudes are not present at the leafs of the state tree (Fig.~\ref{fig:statetree2}), the nodes at the levels above, whose children are absent, are not constructed (Fig.~\ref{fig:statetree3}). This eliminates some branches from the tree.

As seen in the previous section (Eq.~\ref{eq:ampencoded}), the size of a top-down sub-state is defined by the parameter $s$ (split), which indicates the number of qubits of the sub-state and, therefore, the number of amplitudes $2^s$ that it can encode. The number of sub-states depends on two parameters, $s$ (split) and $n$ (total number of qubits of the complete state, which also indicates the height of the state tree). With both parameters, the number of sub-states is $2^n/2^s=2^{n-s}$. Therefore, each block of $2^s$ amplitudes from the input vector makes up a sub-state and that for this sparse version if the block is empty (zero amplitudes) the sub-state is not necessary for the construction of the complete state (Fig.~\ref{fig:statetree3}). If a sub-state has all nodes eliminated, it is no longer needed nor represented in the quantum circuit, reducing the necessary space (qubits). Another consequence of eliminating a sub-state is that the second stage of the algorithm (bottom-up) will combine a smaller number of states (Eq.~\ref{eq:dccombine_children}), therefore reducing the number of CSWAP operators and the total depth of the circuit.

The main difference between the dense algorithm (Pseudocode~\ref{alg:bidirectional}) and the sparse one (Pseudocode~\ref{alg:sparse_bidirectional}) is the construction of the state tree. Function \initbidirectional differs from function \\ \initsparsebidirectional only by Line~\ref{line:sparse_bidirectional_decomposition}, replacing the \statedecomposition function call with \sparsestatedecomposition. In the latter function, the main change that enables sparsity is the introduction of Line~\ref{line:ampright} $if$ conditional statement. This statement has three possible outcomes. The first one (Line~\ref{line:amprightleft}) occurs when two consecutive nodes of even and odd index are present, being identical to the dense case (Pseudocode.~\ref{alg:state_decomposition}). The other two outcomes, first (Line~\ref{line:ampright}) and third (Line~\ref{line:ampleft}) conditions, occur when either of the pair nodes is absent and assumed to have zero amplitude. If the complete pair is absent, none of the statement outcomes are met, then the pair is ignored, and the respective parent node is not created. Node indices are guaranteed to be preserved by Line~\ref{line:sparsenodeindex} of Pseudocode~\ref{alg:sparse_state_decomposition}.

\section{Conclusion}
\label{sec:conclusion}

Existing state preparation methods, such as top-down and bottom-up approaches, require at least one quantum circuit resource between depth and width to grow linearly with the problem size. The BDSP algorithm presented in this work provides a general framework for configuring the trade-off between these resources that can be useful to manage them on NISQ devices. Looking at the state preparation algorithms as a walk on the state tree (see Section~\ref{sec:tree}), the BDSP algorithm constitutes a systematic way to walk in two opposite directions.
Previous methods are based on walking only in one direction. The bidirectional algorithm comes with a free parameter $s\in\lbrack 1, n\rbrack$ that determines the balance between the top-down and the bottom-up approaches. At two extreme cases of setting $s=n$ and $s=1$, the top-down and the bottom-up approaches are respectively recovered. At the equilibrium point $s=\lceil n/2\rceil$, quadratic reduction in both quantum circuit depth and width can be achieved. The configuration parameter can be viewed as a hyperparameter that can tune circuit sizes and the number of CNOT gates according to the compound of application and hardware properties.
The BDSP method is validated and demonstrated through experiments performed on five real quantum devices. The experiments behaved as expected, according to the asymptotic and numerical analyses of the circuit complexity. 

A possible future work is to investigate whether the quantum circuit cost of the DCSP part can be futher reduced. Note that the structure of CSWAP operations in the DCSP step only depends on the dimensionality of the dataset $N$. Hence, the CSWAP operations can be interpreted as a single layer of fixed operation. Decomposing this fixed operation more efficiently than the naive application of CSWAP gates would achieve further reduction in the quantum circuit depth. 

\section*{Acknowledgments}
This work is based upon research supported by CNPq (Grant No. 308730/2018-6, No. 306727/2017-0, No. 409415/2018-9 and No. 421849/2016-9), CAPES – Finance Code 001, FACEPE (Grant No. IBPG-0834-1.03/19), National Research Foundation of Korea (Grant No. 2019R1I1A1A01050161 and No. 2021M3H3A1038085), the South African Research Chair Initiative, Grant No. 64812, of the Department of Science and Innovation and the National Research Foundation (NRF). Support of the NICIS (National Integrated Cyberinfrastructure System) e-research grant QICSA is kindly acknowledged.
We acknowledge the use of IBM Quantum services for this work. The views expressed are those of the authors, and do not reflect the official policy or position of IBM or the IBM Quantum team.

\section*{Competing interests}
The authors declare no competing interests.

\section*{Data availability}
The site \url{https://www.cin.ufpe.br/~ajsilva/qclib} contains all the data and software generated during the current study.

\bibliographystyle{elsarticle-num} 
\bibliography{cas-refs}

\newpage

\appendix

\section{Pseudocode}
\label{apx:pseudocode}

Pseudocodes 1 to 5 expresses algorithms 1 to 5. Pseudocodes \ref{alg:state_decomposition} and \ref{alg:angle_tree} construct the tree representations of the state preparation algorithms, namely the state tree and the angle tree, as described in Section \ref{sec:tree}. Pseudocodes \ref{alg:top_down} and \ref{alg:bottom_up}, which algorithms are explained in sections \ref{sec:top_down} and \ref{sec:bottom_up}, build quantum circuits using top-down and bottom-up approaches for encoding a complex input vector into the amplitudes of a quantum state. Pseudocode \ref{alg:bidirectional} employs pseudocodes 1 to 4 and expresses the bidirectional state preparation algorithm (Sec.~\ref{sec:bidirectional}, Alg.~\ref{alg:bidirectional2}).

Lines \ref{line:bidirectional_stage1} and \ref{line:bidirectional_stage2} of Pseudocode \ref{alg:bidirectional} indicate the two stages of the BDSP algorithm. Line~\ref{line:bidirectional_stage1} at function \topdowntreewalk performs the first stage preparing the sub-states expected by the next stage, equivalent to what would be generated by bottom-up DCSP up to the tree split, but with the absence of ancilla due to the top-down approach. Line~\ref{line:bidirectional_stage2} at function \bottomuptreewalk performs the second stage, starting at level $s+1$ with the sub-states initialized by the previous stage. Line \ref{line:splitstates} at function \topdowntreewalk configures the recurrence so that at split level $s$ it divides the angle tree into $2^{n-s}$ (number of nodes at split level $s$) sub-trees of height $s$, loading all these sub-trees concurrently using the top-down strategy. Lines \ref{line:rotry} and \ref{line:rotrz} of function \bottomuptreewalk initialize $2^{n-s}-1$ qubits exclusive to the second stage with values $R_y(\alpha_{j,v})$ and $R_z(\lambda_{j,v})$. Then function \cswaps combine the states through CSWAP gates controlled by the nodes above level $s$. With the tree described in Fig.~\ref{fig:angletree} and $s=2$, the bidirectional procedure (Pseudocode~\ref{alg:bidirectional}) generates the circuit present in Fig.~\ref{fig:bddetail}.

\begin{pseudocode}
    \SetKwInOut{Input}{input}\SetKwInOut{Output}{output}
    
    \SetKwProg{Fn}{}{:}{}
        \Fn{\statedecomposition{nqubits, data}}
        { \label{func:state_decomposition} 
            \Input{Number of qubits (nqubits) required to generate a state with the same length as the data vector ($2^{nqubits}$).}
            \Input{A list (data) representing the state to be decomposed, with exactly $2^{nqubits}$ pairs (index, amplitude).}
            \Output{Root of the state tree.}
            
            \tcp{Initialize an auxiliary vector \textnormal{new\_nodes} with data vector amplitudes}
            
            new\_nodes = [] \\
            \For{$k\leftarrow 0$ \KwTo $length(data)-1$} 
        	{
        	    node.index = data[k].index \\
        	    node.level = nqubits \\
        	    node.amplitude = data[k].amplitude \\
        		new\_nodes[k] = node
        	}
            
            \tcp{Build the state tree}
            
            \For{$level \leftarrow nqubits$ \KwTo $1$ \KwStep $-1$} 
            { 
                nodes = new\_nodes \\
                new\_nodes = [] \\
                \For{$k\leftarrow 0$ \KwTo $length(nodes)-1$ \KwStep $2$} 
                {
                    mag = $\sqrt{\abs{\textnormal{nodes[k].amplitude}}^2 + \abs{\textnormal{nodes[k+1].amplitude}}^2 }$ \\
                    arg = $( \angle \textnormal{nodes[k].amplitude} + \angle \textnormal{nodes[k+1].amplitude} ) / 2$ \\
                    node.index = nodes[k].index // 2 \\
                    node.level = level \\
                    node.amplitude = $\textnormal{mag} \times \exp(1\textnormal{j} \times \textnormal{arg})$ \\
                    node.left = nodes[k] \\
                    node.right = nodes[k+1] \\
                    new\_nodes[k//2] = node
                }
            }
            
            \KwRet new\_nodes[0] \tcp*{return tree root}
        }
    \caption{Generate a state tree by the decomposition of an amplitude input vector}
    \label{alg:state_decomposition}
\end{pseudocode}

\begin{pseudocode}
    \SetKwInOut{Input}{input}\SetKwInOut{Output}{output}
    \SetKwProg{Fn}{}{:}{}
        \Fn{\angletree{state\_tree}}
        { \label{func:angle_tree} 
            \Input{An output of \statedecomposition function ($state\_tree$).}
            \Output{Tree with angles that will be used to perform the state preparation.}
            \BlankLine
            angle\_y, angle\_z = 0 \label{line:leftchild} \\
            
            \If{$state\_tree.right \ne \Null$}{
                amp = 0 \\
                \If{$state\_tree.amplitude \ne 0$}{
                    amp = state\_tree.right.amplitude / state\_tree.amplitude \label{line:angletreeamp}
                }
                angle\_y = $2 \arcsin(\abs{\textnormal{amp}})$ \\
                angle\_z = $2 \angle \textnormal{amp}$ \\
            }
            node.index = state\_tree.index \\
            node.level = state\_tree.level \\
            node.angle\_y = angle\_y \\
            node.angle\_z = angle\_z \\
            \If{$state\_tree.left \ne \Null \And !\isleaf(state\_tree.left)$}{
                node.left = \angletree{state\_tree.left} \\
            }
            \If{$state\_tree.right \ne \Null \And !\isleaf(state\_tree.right)$}{
                node.right = \angletree{state\_tree.right}
            }
            
            \KwRet node
        }
        
    \caption{Generate a angle tree that will be used to perform the state preparation}
    \label{alg:angle_tree}
\end{pseudocode}

\begin{pseudocode}
    \SetKwInOut{Input}{input}\SetKwInOut{Output}{output}
    \SetKwFunction{uniformlycontrolledrotation}{uniformly\_controlled\_rotation}
    \SetKwFunction{children}{children}
    \SetKwFunction{append}{append}
    
    \SetKwProg{Fn}{}{:}{}

        \Fn{\topdowntreewalk{angle\_tree, circuit, start\_level, control\_nodes=\textnormal{null}, target\_nodes=\textnormal{null}}}
        { \label{func:top_down_tree_walk} 
            \Input{An output of \angletree function ($angle\_tree$).}
            \Input{A quantum circuit to apply the top-down encoding ($circuit$).}
            \Input{The tree level to start the walk ($start\_level$).}
            \Input{Used in the recursive calls ($control\_nodes$).}
            \Input{Used in the recursive calls ($target\_nodes$).}
            \Output{$circuit$ after the application of the top-down encoding.}
            
            \BlankLine
            
            \If{$angle\_tree \ne \Null$}{
                \eIf{$angle\_tree.level < start\_level$ \label{line:splitstates}}{
                    \topdowntreewalk{angle\_tree.left, circuit, start\_level} \\
                    \topdowntreewalk{angle\_tree.right, circuit, start\_level}
                }{
                    angle\_tree.qubit = \addqubit{circuit} \label{line:topdown_addqubit} \\
                    
                    \If{$target\_nodes == \Null$}{
                        control\_nodes = [] \tcp*{initialize the controls list}
                        target\_nodes[0] = angle\_tree \tcp*{start by the sub-tree root}
                    }
                    \uniformlycontrolledrotation{circuit, control\_nodes, target\_nodes} \\
                    \append{control\_nodes, angle\_tree} \tcp*{add curr. node to the controls list}
                    target\_nodes = \children{target\_nodes} \tcp*{all the nodes in the next level}
                   
                    \eIf{$angle\_tree.left \ne \Null$}{
                        \topdowntreewalk{angle\_tree.left, circuit, start\_level, control\_nodes, target\_nodes}
                    }{
                        \topdowntreewalk{angle\_tree.left, circuit, start\_level, control\_nodes, target\_nodes}
                    }
                }
            }
            
        }
        
        \Fn{\inittopdown{circuit, state}}
        { \label{func:initialize_top_down} 
            nqubits = $\log_2(\length(state))$ \\
            state\_tree = \statedecomposition{nqubits, state}\\
            angle\_tree = \angletree{state\_tree}
            \BlankLine
            \topdowntreewalk{angle\_tree, circuit, 0}
            \BlankLine
            output\_nodes = \leftview{angle\_tree} \\
            \For{$k\leftarrow 0$ \KwTo $nqubits-1$} 
        	{
        	    output\_qubits[k] = output\_nodes[k].qubit
        	}
            \KwRet output\_qubits
        }
    
    \caption{Construct a circuit that perform a top-down state preparation for the input vector $state$. The intended quantum state is encoded on the $output\_qubits$.}
    \label{alg:top_down}
\end{pseudocode}

\begin{pseudocode}
    \SetKwInOut{Input}{input}\SetKwInOut{Output}{output}
    
    \SetKwProg{Fn}{}{:}{}
        \Fn{\cswaps{angle\_tree, circuit}}
        { \label{func:cswaps} 
            \Input{An output of \angletree function ($angle\_tree$).}
            \Input{A quantum circuit to apply the cswaps ($circuit$).}
            \Output{$circuit$ after the application of the cswaps.}
            \BlankLine
            left = angle\_tree.left\\
            right = angle\_tree.right\\
            \While{$left \ne \Null \And right \ne \Null$}
            {
                circuit.cswap($angle\_tree.qubit, left.qubit, right.qubit$)\\
                left = left.left\\
                right = right.left
            }
        }
        
        \Fn{\bottomuptreewalk{state\_tree, circuit, start\_level}}
        { \label{func:bottom_up_tree_walk} 
            \Input{An output of \statedecomposition function ($state\_tree$).}
            \Input{A quantum circuit to apply the bottom-up encoding ($circuit$).}
            \Input{The tree level to start the bottom-up walk ($start\_level$).}
            \Output{$circuit$ after the application of the bottom-up encoding.}
            
            \BlankLine
            
            \If{$angle\_tree \ne \Null \And angle\_tree.level < start\_level$}{
                angle\_tree.qubit = \addqubit{circuit} \label{line:bottomup_addqubit} \\
                circuit.ry($angle\_tree.angle\_y, angle\_tree.qubit$) \label{line:rotry}\\
                circuit.rz($angle\_tree.angle\_z, angle\_tree.qubit$) \label{line:rotrz}\\
                \bottomuptreewalk{angle\_tree.left, circuit, start\_level}\\
                \bottomuptreewalk{angle\_tree.right, circuit, start\_level}\\
                
                \cswaps{angle\_tree, circuit}
            }
        }
        
        \Fn{\initbottomup{circuit, state}}
        { \label{func:initialize_bottom_up}
            nqubits = $\log_2(\length(state))$ \\
            
            state\_tree = \statedecomposition{nqubits, state}\\
            angle\_tree = \angletree{state\_tree}
            \BlankLine
            \bottomuptreewalk{angle\_tree, circuit, nqubits}
            \BlankLine
            output\_nodes = \leftview{angle\_tree} \\
            \For{$k\leftarrow 0$ \KwTo $nqubits-1$} 
        	{
        	    output\_qubits[k] = output\_nodes[k].qubit
        	}
            \KwRet output\_qubits
        }
        
    \caption{Construct a circuit that perform a bottom-up state preparation for the input vector $state$. The intended quantum state is encoded on the $output\_qubits$.}
    \label{alg:bottom_up}
\end{pseudocode}

\begin{pseudocode}
    \SetKwInOut{Input}{input}\SetKwInOut{Output}{output}
    
    \SetKwProg{Fn}{}{:}{}
        \Fn{\initbidirectional{$circuit, state, split$}}
        { \label{func:initialize_bidirectional}
            nqubits = $\log_2(\length(state))$ \\
            state\_tree = \statedecomposition{nqubits, state} \\
            angle\_tree = \angletree{state\_tree}
            \BlankLine
            \topdowntreewalk{angle\_tree, circuit, nqubits$-$split} \label{line:bidirectional_stage1} \tcp*{stage 1}
            \bottomuptreewalk{angle\_tree, circuit, nqubits$-$split} \label{line:bidirectional_stage2} \tcp*{stage 2}
            \BlankLine
            output\_nodes = \leftview{angle\_tree} \\
            \For{$k\leftarrow 0$ \KwTo $nqubits-1$} 
        	{
        	    output\_qubits[k] = output\_nodes[k].qubit
        	}
            \KwRet output\_qubits
        }
        
    \caption{Construct a circuit that perform a bidirectional state preparation for the input vector $state$. The intended quantum state is encoded on the $output\_qubits$.}
    \label{alg:bidirectional}
\end{pseudocode}

\begin{pseudocode}
    \SetKwInOut{Input}{input}\SetKwInOut{Output}{output}
    
    \SetKwProg{Fn}{}{:}{}
        \Fn{\initsparsebidirectional{$circuit, state, split$}}
        { \label{func:initialize_sparse_bidirectional}
            nqubits = $\log_2(\length(state))$ \\
            state\_tree = \sparsestatedecomposition{nqubits, state} \label{line:sparse_bidirectional_decomposition} \\
            angle\_tree = \angletree{state\_tree}
            \BlankLine
            \topdowntreewalk{angle\_tree, circuit, nqubits$-$split} \label{line:sparse_bidirectional_stage1} \tcp*{stage 1}
            \bottomuptreewalk{angle\_tree, circuit, nqubits$-$split} \label{line:sparse_bidirectional_stage2} \tcp*{stage 2}
            \BlankLine
            output\_nodes = \leftview{angle\_tree} \\
            \For{$k\leftarrow 0$ \KwTo $nqubits-1$} 
        	{
        	    output\_qubits[k] = output\_nodes[k].qubit
        	}
            \KwRet output\_qubits
        }
        
    \caption{Construct a circuit that perform a bidirectional state preparation for the sparse input vector $state$. The intended quantum state is encoded on the $output\_qubits$.
    } \label{alg:sparse_bidirectional}
\end{pseudocode}

\begin{pseudocode}
    \SetKwInOut{Input}{input}\SetKwInOut{Output}{output}
    
    \SetKwProg{Fn}{}{:}{}
        \Fn{\sparsestatedecomposition{nqubits, data}}
        { \label{func:sparse_state_decomposition} 
            \Input{Number of qubits (nqubits) required to generate a state with the same length as the data vector ($2^{nqubits}$).}
            \Input{A list (data) representing the state to be decomposed, with exactly $2^{nqubits}$ pairs (index, amplitude).}
            \Output{Root of the state tree.}
            
            \tcp{Initialize an auxiliary vector \textnormal{new\_nodes} with data vector amplitudes}
            
            new\_nodes = [] \\
            \For{$k\leftarrow 0$ \KwTo $length(data)-1$} 
        	{
        	    node.index = data[k].index \\
        	    node.level = nqubits \\
        	    node.amplitude = data[k].amplitude \\
        		new\_nodes[k] = node
        	}
            
            \tcp{Build the state tree}
            
            \For{$level \leftarrow nqubits$ \KwTo $1$ \KwStep $-1$} 
            { 
                nodes = new\_nodes \\
                new\_nodes = [] \\
                \For{$k\leftarrow 0$ \KwTo $length(nodes)-1$ \KwStep $2$} 
                {
                    node.index = nodes[k].index // 2 \label{line:sparsenodeindex} \\
                    node.level = level \\
                    \BlankLine
                    mag = $\abs{nodes[k].amplitude}$ \\
                    arg = $(\angle nodes[k].amplitude) / 2$ \\
                    \uIf{$nodes[k].index \% 2 == 1$ \label{line:ampright}}{
                        node.left = \Null \\
                        node.right = nodes[k] \\
                        k = k - 1
                    }
                    \uElseIf{$(k + 1) < \length(nodes) \And nodes[k + 1].index == nodes[k].index + 1$ \label{line:amprightleft}}{
                        mag = $\sqrt{\abs{\textnormal{nodes[k].amplitude}}^2 + \abs{\textnormal{nodes[k+1].amplitude}}^2 }$ \\
                        arg = $( \angle \textnormal{nodes[k].amplitude} + \angle \textnormal{nodes[k+1].amplitude} ) / 2$ \\
                        node.left = nodes[k] \\
                        node.right = nodes[k+1] \\
                    }
                    \Else{
                        \label{line:ampleft}
                        node.left = nodes[k]\\
                        node.right = \Null \\
                        k = k - 1
                    }
                    node.amplitude = $\textnormal{mag} \times \exp(1\textnormal{j} \times \textnormal{arg})$ \\
                    \BlankLine
                    new\_nodes.append(node)
                }
            }
            
            \KwRet new\_nodes[0] \tcp*{return tree root}
        }
    \caption{Generate a sparse state tree by the decomposition of an sparse amplitude input vector
    } \label{alg:sparse_state_decomposition}
\end{pseudocode}

\end{document}